\documentclass[sigconf]{acmart}
\usepackage{graphicx}
\usepackage{textcomp}
\usepackage{xcolor}

\usepackage{tikz}
\usepackage{pgfplots}
\usepackage{pdfpages}
\usepackage{pgfplotstable}
\usetikzlibrary{patterns,automata,positioning,spy,arrows,decorations, matrix}
\usepgfplotslibrary{groupplots, statistics}
\usepackage{siunitx}
\usepackage{caption,subcaption}
\usepackage{comment}
\usepackage{csvsimple}
\usepackage{algorithmic}
\usepackage[algo2e,ruled,vlined]{algorithm2e}
\usepackage{graphicx}
\usepackage{calc}
\usepackage{multirow}
\usepackage{array}
\usepackage{booktabs}
\usepackage{enumitem}
\usepackage{url}
\usepackage{listings}
\usepackage{dblfloatfix}
\pgfplotsset{compat=1.16}
\newcommand{\workname}{MAPA}
\newcommand{\workfullname}{Multi-Accelerator Pattern Allocation}

\newcommand*{\boxplot}[6]{%
  \addplot+[
    line width=.2mm,
    black,
    boxplot prepared={
      lower whisker={#5},
      lower quartile={#3},
      median={#2},
      upper quartile={#4},
      upper whisker={#6},
    }
  ]
  coordinates{};
}

\newcommand\NOTES{}
\ifdefined\NOTES
\newcommand{\daniel}[1]{\textcolor{red}{\small \textsf{DW: #1}}}
\newcommand{\kiran}[1]{\textcolor{brown}{Kiran: \em #1 }}
\newcommand{\revision}[1]{\textcolor{blue}{{#1}}}

\fi
\unless\ifdefined\NOTES
\newcommand{\daniel}[1]{}
\newcommand{\kiran}[1]{}
\newcommand{\revision}[1]{}
\fi

\def\BibTeX{{\rm B\kern-.05em{\sc i\kern-.025em b}\kern-.08em
    T\kern-.1667em\lower.7ex\hbox{E}\kern-.125emX}}
    
\setcopyright{rightsretained} 

\begin{document}

\title{MAPA: Multi-Accelerator Pattern Allocation Policy for Multi-Tenant GPU Servers}

\author{Kiran Ranganath}
\affiliation{
Department of Electrical and Computer Engineering\\
University of California Riverside
\country{CA, USA}
}
\email{krang006@ucr.edu}

\author{Joshua D. Suetterlein}
\affiliation{
High-Performance Computing Group\\
Pacific Northwest National Lab\\
\country{WA, USA}
}
\email{joshua.suetterlein@pnnl.gov}

\author{Joseph B. Manzano}
\affiliation{
High-Performance Computing Group\\
Pacific Northwest National Lab\\
\country{WA, USA}
}
\email{joseph.manzano@pnnl.gov}

\author{Shuaiwen Leon Song}
\affiliation{
Future System Architecture Lab\\
School of Computer Science\\
University of Sydney\\
\country{Sydney, Australia}
}
\email{shuaiwen.song@sydney.edu.au}

\author{Daniel Wong}
\affiliation{
Department of Electrical and Computer Engineering\\
University of California Riverside\\
\country{CA, USA}
}
\email{danwong@ucr.edu}

\begin{abstract}
Multi-accelerator servers are increasingly being deployed in shared multi-tenant environments (such as in cloud data centers) in order to meet the demands of large-scale compute-intensive workloads. In addition, these accelerators are increasingly being inter-connected in complex topologies and workloads are exhibiting a wider variety of inter-accelerator communication patterns. However, existing allocation policies are ill-suited for these emerging use-cases.
Specifically, this work identifies that multi-accelerator workloads are commonly \textit{fragmented} leading to reduced bandwidth and increased latency for inter-accelerator communication. 

We propose Multi-Accelerator Pattern Allocation (\workname), a graph pattern mining approach towards providing generalized allocation support for allocating multi-accelerator workloads on multi-accelerator servers.
We demonstrate that \workname~ is able to improve the execution time of multi-accelerator workloads and that \workname~ is able to provide generalized benefits across various accelerator topologies. Finally, we demonstrate a speedup of 12.4\% for 75th percentile of jobs with the worst case execution time reduced by up to 35\% against baseline policy using MAPA.

\end{abstract}

\copyrightyear{2021} 
\acmYear{2021} 
\acmConference[SC '21]{The International Conference for High Performance Computing, Networking, Storage and Analysis}{November 14--19, 2021}{St. Louis, MO, USA}
\acmBooktitle{The International Conference for High Performance Computing, Networking, Storage and Analysis (SC '21), November 14--19, 2021, St. Louis, MO, USA}
\acmDOI{10.1145/3458817.3480853}
\acmISBN{978-1-4503-8442-1/21/11}

\settopmatter{printfolios=true}
\maketitle

\section{Introduction}
\label{sec:introduction}

The never ending demand for faster computation from data intensive workloads has driven the growth for multi-accelerator servers. Systems equipped with accelerators, such as General Purpose Processing in Graphical Processing Units (GPGPUs) and Tensor Processing Units (TPU)~\cite{tpu} are increasingly being deployed in shared environments, such as Cloud, Enterprise, and High-Performance Computing (HPC). These systems are increasingly modular with many accelerators within a single server.

\begin{figure}[!htb]
    \begin{subfigure}{0.3\linewidth}
        \centering
        \includegraphics[width=0.57\linewidth]{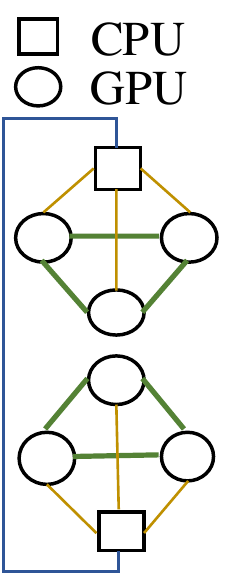}
        \caption{Summit V100}
        \label{fig:summit}
    \end{subfigure}
    \begin{subfigure}{0.3\linewidth}
        \centering
        \includegraphics[width=0.66\linewidth]{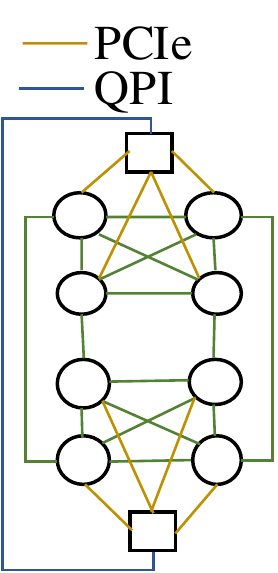}
        \caption{DGX-1 P100}
        \label{fig:dgx_p100}
    \end{subfigure}
    \begin{subfigure}{0.3\linewidth}
        \centering
        \includegraphics[width=\linewidth]{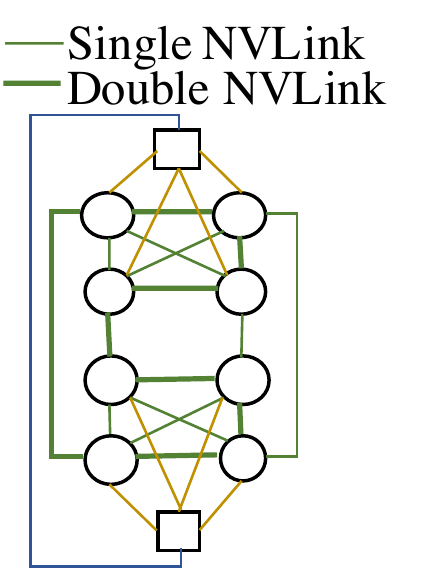}
        \caption{DGX-1 V100}
        \label{fig:dgx_v100}
    \end{subfigure}
    \vspace{-3mm}
    \caption{Emerging multi-GPU accelerator topologies are increasingly heterogeneous.}
    \label{fig:multi_gpu_topologies_in_use}
    \vspace{-4mm}
\end{figure}

As software and hardware becomes more complex and heterogeneous, new challenges have emerged in software-hardware stack. 

Two major challenges of modern large-scale systems are the need for faster collective communication operations~\cite{wang2019blink, ranganath2019speeding} and topology-aware scheduling~\cite{amaral2017topology, gandiva}. Recent works like topology-aware scheduling~\cite{amaral2017topology} and Gandiva~\cite{gandiva} have motivated the importance of optimal placements to improve performance of Machine Learning (ML) workloads within multi-GPU environments by efficiently utilizing inter-accelerator interconnection link. Systems such as Nvidia's DGX-V100, Facebook's Big-Basin~\cite{big-basin-opencompute}, and  Amazon's P3DN~\cite{amazon-p3dn} have accelerators connected with many different types of interconnection links.

\begin{figure*}[!ht]
    \begin{tikzpicture}
        \begin{groupplot}[
            group style = {
                group size = 2 by 1,
                horizontal sep = 1.5cm
            }
        ]
        \nextgroupplot [
            align=center,
            title={(a) Bandwidth characterization},
            xmode=log, mark size=1pt,
            xlabel=$Data~Size~(Bytes)$, ylabel=$Bandwidth~(GB/s)$, legend pos=north west,
            width = \columnwidth, height = 3.5cm,
            title style={at={(0.5,-0.7)},anchor=north,},
            legend style={legend columns=1, nodes={scale=0.7}},
            ]
            \addplot table [x=size, y=NV2S, col sep=space] {data/bw-data.csv};
            \addplot table [x=size, y=NV2D, col sep=space] {data/bw-data.csv};
            \addplot table [x=size, y=PCIe, col sep=space] {data/bw-data.csv};
            \legend{NV2-Single, NV2-Double, PCIe}
        
        \nextgroupplot [
                align=center,
                title={(b) Speedup with different links},
                title style={at={(0.5,-0.7)},anchor=north,},
                ylabel=$Speedup$, xlabel=$Network$,
                symbolic x coords={VGG-16, Resnet, AlexNet, Inception, CaffeNet, GoogleNet},
                xtick=data, legend pos=north east,
                width = \columnwidth, height = 3.5cm,
                tick label style={font=\small},
                bar width=0.2cm,
                enlarge x limits=0.08,
                ybar=0pt,
                legend columns=3,
                ymin = 0, ymax=3.7,
                legend style={legend columns=2, nodes={scale=0.7}},
            ]
            \addplot [
                    ybar,
                    fill=blue!40!white
                ] table [x=nets, y=NV2D, col sep=space] {data/net-mot-data.csv};
            \addplot [
                    ybar,
                    fill=red!40!white
                ] table [x=nets, y=NV2S, col sep=space] {data/net-mot-data.csv};
            \addplot [
                    ybar,
                    fill=black!40!white
                ] table [x=nets, y=PCIe, col sep=space] {data/net-mot-data.csv};
            \legend{NV2-Double, NV2-Single, PCIe}
        \end{groupplot}
    \end{tikzpicture}
    \caption{Nvidia's multi-GPU systems exhibit a variety of interconnects. This figure shows the various links available in DGX-1 Volta. These different links have significantly different bandwidth as well as impact on applications such as CNN training. where NV2-Single and NV2-Double are Single and Double NVLink-v2 links respectively.}
    \label{fig:motivation_charts}
\end{figure*}
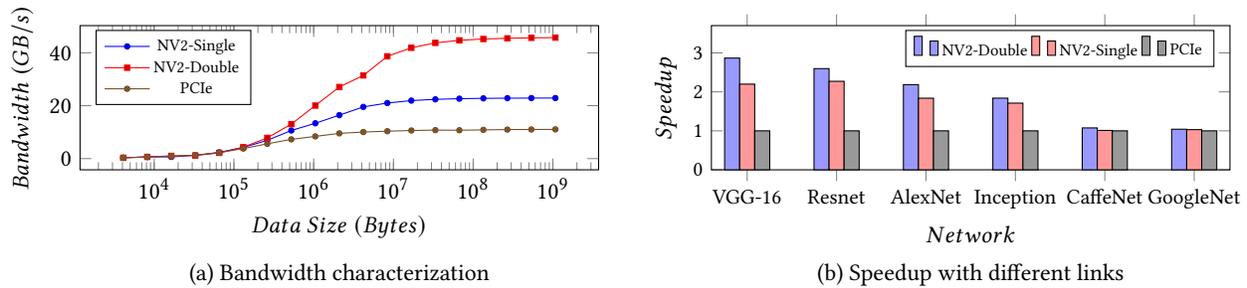

In this work, we focus on the challenge of inefficient job allocation in a multi-accelerator environment.  These sub-par allocations can lead to significant slowdown in execution time. These challenges are most prominent in architectures with high heterogeneity in their inter-accelerator inter-connect network (i.e., different number of links with different bandwidths, non uniform network accesses, etc.) such as NVIDIA\rq s DGX-1(Figure~\ref{fig:dgx_p100}),  Facebook's Big-Basin systems~\cite{big-basin-opencompute}, Amazon's P3DN~\cite{amazon-p3dn} and DGX-1-V (Figure~\ref{fig:dgx_v100}). Even designs with constant access latency such as the DGX-2 exhibit NUMA effects~\cite{li2019} which can lead to allocation inefficiencies. 
Furthermore, new accelerator designs such as  TPUs~\cite{tpu} and multi-chip accelerators~\cite{chiplet-based-systems} can further fuel the adoption of heterogeneous multi-accelerator designs. As the number of accelerators continues to grow, a smarter job scheduler and resource allocator is needed to fully utilize the underlying hardware and  handle the increasing complexity of multi-accelerator workloads. 

To this end, we propose a graph pattern matching-based allocation solution called \workfullname (\workname) to address problems with allocation of multi-accelerator workloads in multi-accelerator environments. \workname\ aims to  provide a generic framework applicable to any multi-accelerator environment.

The contributions of this paper are the following:
\begin{itemize}[noitemsep,topsep=0pt,parsep=0pt,partopsep=0pt]
    \item Performance analysis of increasingly heterogeneous accelerator communication links (i.e., PCIe, NVLink) to motivate the need for hardware topology-aware allocation policies.
    \item \workname, a graph pattern matching approach for scheduling multi-accelerator workloads on multi-accelerator systems. 
    \item Novel metrics to score matching patterns and predict the effective bandwidth of an allocation. 
    \item Evaluation of \workname~with machine learning training workloads on real-world multi-GPU server.
    \item Exploration of \workname~ on novel hardware topologies at larger scale and complex non-uniform topologies.
\end{itemize}

\section{Motivation}
\label{sec:motivation}

Increasingly more popular cloud \cite{aws-gpu, google-cloud, big-basin-opencompute, dgx1Spec, dgx2Spec} and modern HPC systems \cite{summit} are accelerator based, and are used to train and to deploy complex machine learning workloads across many different fields from proteomics to self driving vehicles.  While these systems primarily employ GPUs, in the future, systems are expected to take advantage of other types of accelerators such as FPGAs or TPUs~\cite{tpu}.  The following describes some of the challenges posed by these multi-accelerator architectures.

\subsection{Modern Multi-Accelerator Systems}
\label{subsec:modern_multi_accelerator_systems}
\textit{Characterizing Accelerator Interconnects. }
Modern multi-GPU servers exhibit a wide range of capabilities when dealing with inter-GPU communication. Table~\ref{table:peak_bandwidth} lists the types of links used to connect accelerators in these systems and their respective bandwidths. 

\begin{table}[!ht]
\small
    \centering
    \begin{tabular}{|c|c|}
        \hline
         \textbf{Link} & \textbf{Bandwidth (GBps)}\\
         \hline
         Single NVlink-v1 & 20\\
         \hline
         Single NVlink-v2 & 25\\
         \hline
         Double NVlink-v2 & 50\\
         \hline
         16-lanes PCIe Gen 3~\cite{nvidia2017v100} & 12\\
         \hline
    \end{tabular}
    \caption{Peak Bandwidths per link}
    \label{table:peak_bandwidth}
\end{table}

In systems like Big basin~\cite{big-basin-opencompute}, P3DN~\cite{amazon-p3dn}, Summit~\cite{summit}, DGX-1 V100~\cite{dgx2Spec}, and DGX-1 P100~\cite{dgx1Spec}, accelerators are not uniformly connected.  For example, in earlier generations of the DGX systems, communication can be routed through PCIe links in the case that a direct NVLink cannot be found. Furthermore, in the case of DGX-1 with Volta GPUs (a.k.a. DGX1-V100) and Big-basin, there are some accelerators that are connected via double NVLink connections.
Current support for communicating and synchronizing across accelerators includes NVidia Collective Communication Libraries (NCCL)~\cite{nccl}, AMD's Radeon Collective Communication Library (RCCL)~\cite{rccl}, and Baidu All-Reduce~\cite{baiduallreduce}.

\begin{figure}
    \centering
    \begin{tikzpicture}
        \begin{groupplot}[
            group style = {
                group size = 2 by 1,
                horizontal sep = 1.3cm
            }
        ]
        \nextgroupplot
        [
            xlabel=$Year$, ylabel=$HPC~Systems$, ylabel near ticks,
            title={(a) Accelerator-based},
            ymin=0,ymax=220,
            ybar stacked,
            width=0.5\columnwidth, height=3.5cm, 
            symbolic x coords={2017, 2018, 2019, 2020, 2021}, xtick=data,
            tick label style={font=\small},
            title style={at={(0.4,-0.45)},anchor=north,},
            legend style={legend columns=2, nodes={scale=0.6},}, legend pos=north west,
        ] \addplot [fill=black!20!white] table
        [x=Year, y=GPU-based-SC, col sep=space] {data/top500-interconnects.csv};
        \addplot [fill=black!50!white] table [x=Year, y=Other-Acc, col sep=space] {data/top500-interconnects.csv};
        \legend{GPU, Others}
        \nextgroupplot
        [
            xlabel=$Year$, ylabel=$Ratio~(\%)$, ylabel near ticks,
            title={(b) Heterogeneous interconnects},
            width=0.5\columnwidth, height=3.5cm, ymin=0,ymax=105,
            symbolic x coords={2017, 2018, 2019, 2020, 2021}, xtick=data,
            tick label style={font=\small},
            title style={at={(0.5,-0.45)},anchor=north,},
        ] \addplot [color=black,mark=*] table [x=Year, y=Ratio, col sep=space] {data/top500-interconnects.csv};
        \end{groupplot}
    \end{tikzpicture}
    \caption{The number of Top500 supercomputers with accelerators are increasing, with GPUs being the most common. The ratio of these GPU HPC systems with heterogeneous interconnects has increased over time and are now dominant.}
    \label{fig:top_500_interconnect_adoption}
    \vspace{-2mm}
\end{figure}
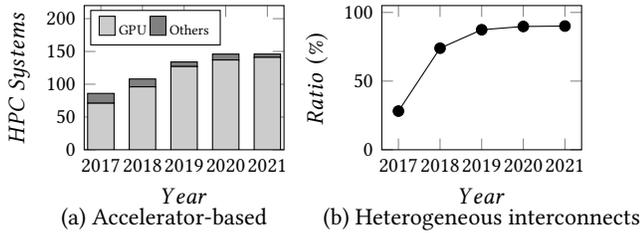

We observe from Figure ~\ref{fig:top_500_interconnect_adoption} (a) that supercomputers are increasingly employing discrete GPUs.
Figure ~\ref{fig:top_500_interconnect_adoption} (b) shows the increased presence of heterogeneous interconnects in such systems. Hence, it is important to identify and explore allocation challenges in such compute environments. Additionally, machine learning-based workloads has recently gained attention in the HPC community, with efforts such as Mesh-tensorflow ~\cite{mesh-tensorflow} and Zero~\cite{zero-trillion-parameters} which aim to improve the scalability and performance of machine learning on supercomputing systems. Furthermore, there exist numerous works that have attempted to utilize machine learning to accelerate various simulation workloads on HPC systems ~\cite{dong2020smart,partee2021using,peterson2019merlin,fox2019learning,brenowitz2020machine,kadupitiya2019machine,kadupitiya2020machine,wang2019defsi}.

In Figure~\ref{fig:motivation_charts}(a), we characterize the communication bandwidth achieved with different links by running the NCCL All-reduce microbenchmark on a DGX-V100 system. This figure demonstrates the peak achievable communication bandwidth of various links across different data transfer sizes. While smaller data transfer sizes achieve lower bandwidth, the relative performance of each link type to each other remains, with double NVLink being the fastest.

In Figure~\ref{fig:motivation_charts}(b), we show the impact of allocation on popular ML training jobs to GPUs connected by these links. 
We obtained this by running Caffe workloads across 2 GPUs to utilize the various interconnects. To utilize double NVLink, single NVLink and PCIe, we allocate to GPUs 1 and 5, 1 and 2, and 1 and 6, respectively. We see that certain networks, such as VGG-16, experience up to 3x execution time speedup using double NVLink compared to PCIe, while other workloads, such as GoogleNet are less impacted. In general, we observe that allocation of high-bandwidth links is critical for workloads with larger data transfers. 

\textit{Multi-tenant Multi-Accelerator Servers. }
It was shown in Philly~\cite{jeon2018multi} and Gandiva~\cite{gandiva} that jobs running in cloud environments often do not use all of the available accelerator resources. Thus to ensure the best return on investment in terms of costs and energy, co-location of jobs might be desirable in order to boost utilization. In fact, co-location has already appeared in modern Nvidia GPUs with the Multi-Instance GPU (MIG)~\cite{nvidia-mig} feature which enables the GPU accelerator to be shared by up to 7 instances.  However, co-location introduces challenges for hard-limit real-time applications, secure applications, or high performance workloads in general. The effects on performance / security for co-locating jobs requires a further in-depth exploration to ensure that the loss in these metrics is acceptable for these applications.

\subsection{Resource fragmentation in multi-tenant servers}
One critical challenge caused by multi-tenant servers is that allocated hardware resources can become \textit{fragmented}, that is, the allocated GPUs can be scattered across the entire topology resulting in the loss of high-bandwidth interconnect available to the workload.
For example, a 3-GPU allocation will experience fragmentation when allocating GPUs 1, 2, and 5 on the DGX-V system shown in Figure~\ref{fig:dgx_v100}. This allocation would require the use of low-bandwidth PCIe that traverses the CPU's QPI interconnect in order to communicate directly between GPU 2 and GPU 5.

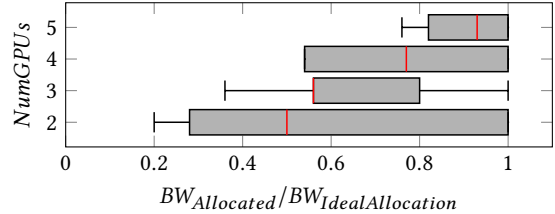
\begin{figure}[!h]
    \begin{tikzpicture}
    \begin{axis}[%
    xmin=0, xmax=1.1,%
    ytick={1,2,3,4},yticklabels={2,3,4,5},%
    boxplot/every median/.style={draw=red},
    ylabel={$NumGPUs$}, xlabel=$BW_{Allocated}/BW_{IdealAllocation}$,
    height=3.5cm, width=0.95\columnwidth,
    every axis plot/.append style={fill, fill opacity=0.3},
    ]
    \boxplot{1}{.5}{.28}{1}{.2}{1}
    \boxplot{2}{.56}{.56}{.8}{.36}{1}
    \boxplot{3}{.77}{.54}{1}{.54}{1}
    \boxplot{4}{.93}{.82}{1}{.76}{1}
    \end{axis}
    \end{tikzpicture}
    \caption{Due to fragmentation of GPU allocations, a large portion of GPU jobs have sub-optimal allocated aggregate bandwidth ($BW_{Allocated}$) compared to the aggregate bandwidth of an ideal allocation ($BW_{IdealAllocation}$).}
    \label{fig:baseline_bw_dist}
\end{figure}

To quantify and highlight this problem, we present Figure~\ref{fig:baseline_bw_dist}. The x-axis shows the \textit{quality of bandwidth allocation} which we quantify as the \textit{aggregate bandwidth of an allocation} ($BW_{Allocated}$) with respect to the \textit{ideal aggregate bandwidth of an ideal allocation} ($BW_{IdealAllocation}$). For example, for a 3-GPU allocation of GPUs 1, 2, and 5, $BW_{Allocated}$ is 87 GBps (1 PCIe, 1 Single NVLink, 1 Double NVLink). The ideal 3-GPU allocation would be GPUs 1, 3, and 4, where $BW_{IdealAllocation}$ is 125 GBps (1 Single NVLink, 2 Double NVLinks).

We ran 100 machine learning training jobs, each utilizing a different number of GPUs (y-axis), on a DGX-V system using the default baseline scheduling in Nvidia Docker where GPUs are assigned to jobs based on the lowest available GPU IDs (see Section~\ref{sec:evaluation} for experimental methodology details). The box-plot shows the distribution of bandwidth allocation quality.

\begin{figure}[!t]
\subfloat[Cumulative distribution of collective communication calls]{
    \centering
    \begin{tikzpicture}
        \begin{axis} [xmode=log, ylabel=$CDF$, width=0.8\linewidth, height=3.5cm,
        legend style={legend columns=1, nodes={scale=0.7}},
        legend pos = outer north east,, xlabel=$Size~(Bytes)$,
        ]
        \addplot [color=blue] table [x=size, y=cdf, col sep=space] {data/network-communication-cdf/alexnet.csv};
        \addplot [color=orange!50!black] table [x=size, y=cdf, col sep=space] {data/network-communication-cdf/inception-v3.csv};
        \addplot [color=green!50!black] table [x=size, y=cdf, col sep=space] {data/network-communication-cdf/vgg-16.csv};
        \addplot [color=black] table [x=size, y=cdf, col sep=space] {data/network-communication-cdf/resnet-50.csv};
        \addplot [color=purple] table [x=size, y=cdf, col sep=space] {data/network-communication-cdf/caffenet.csv};
        \addplot [color=red] table [x=size, y=cdf, col sep=space] {data/network-communication-cdf/googlenet.csv};
        \legend{AlexNet, GoogleNet, VGG, Resnet, Inception, CaffeNet}
        \end{axis}
    \end{tikzpicture}} \hfill
    
\subfloat[Number of communication calls triggered per GPU per iteration and bandwidth sensitivity]{
\small
    \begin{tabular}{|c|>{\centering\arraybackslash}p{3cm}|>{\centering\arraybackslash}p{2cm}|}
    \hline
         \textbf{Network} & \textbf{Communication calls per iter.} & \textbf{Bandwidth Sensitive} \\
         \hline
         AlexNet & 80,001 & Yes \\
         \hline
         Inception-v3 & 2,830,001 & Yes \\
         \hline
         VGG-16 & 160,001 & Yes\\
         \hline
         Resnet-50 & 1,600,001 & Yes\\
         \hline
         CaffeNet & 84,936 & No\\
         \hline
         GoogleNet & 640,001 & No\\
         \hline
    \end{tabular}
    \label{table:bandwidth_sensitivity_matrix}
    }
    \caption{Communication Properties of ML workloads}
    \label{fig:ml_comm_properties}
\end{figure}

\begin{figure}[!t]
    \centering
    \begin{subfigure}[c]{0.495\linewidth}
        \centering
        \includegraphics[width=1.1\linewidth]{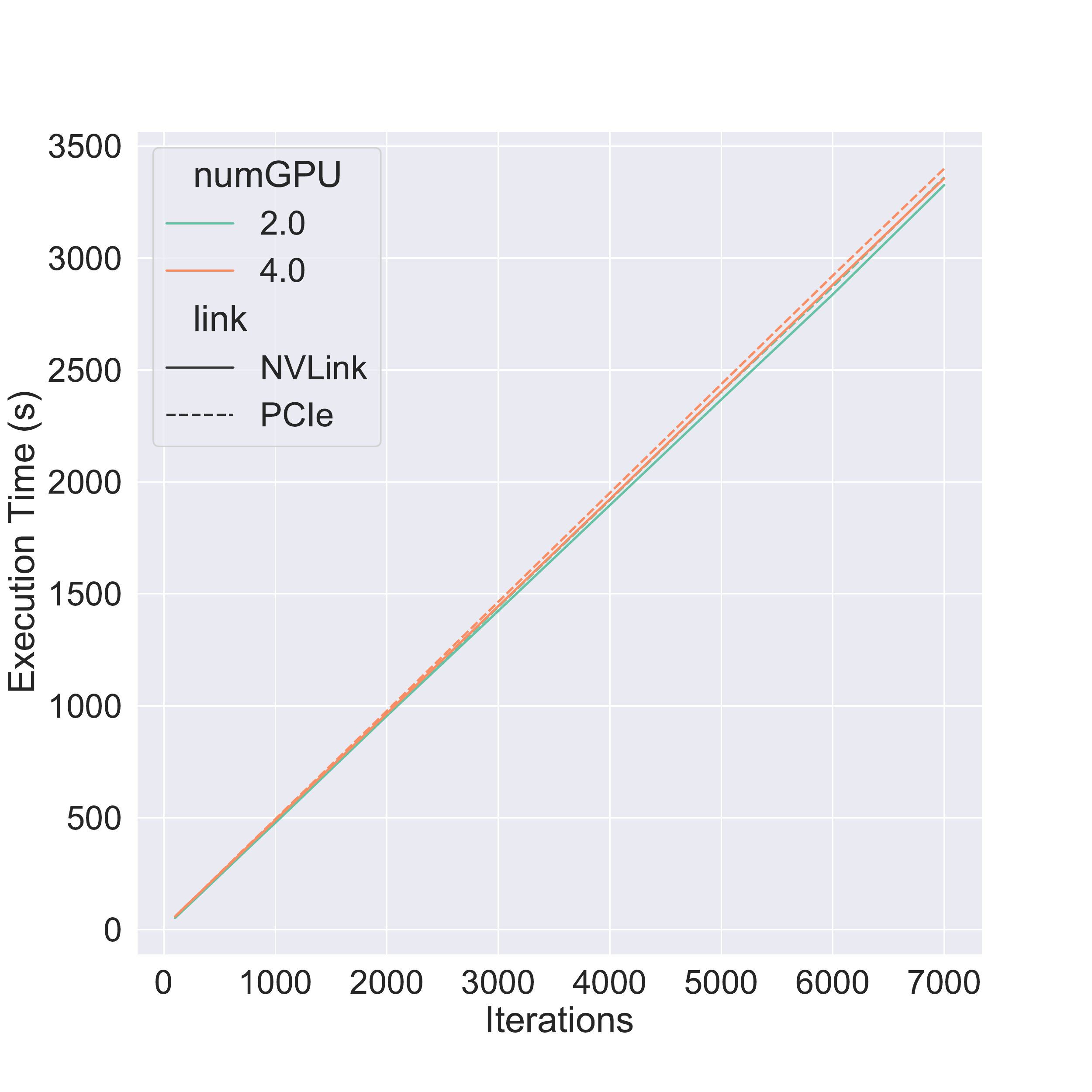}
        \caption{GoogleNet (Insensitive)}
        \label{fig:googlenet_caffe_iter}
    \end{subfigure}
    \hfill
    \begin{subfigure}[c]{0.495\linewidth}
        \centering
        \includegraphics[width=1.1\linewidth]{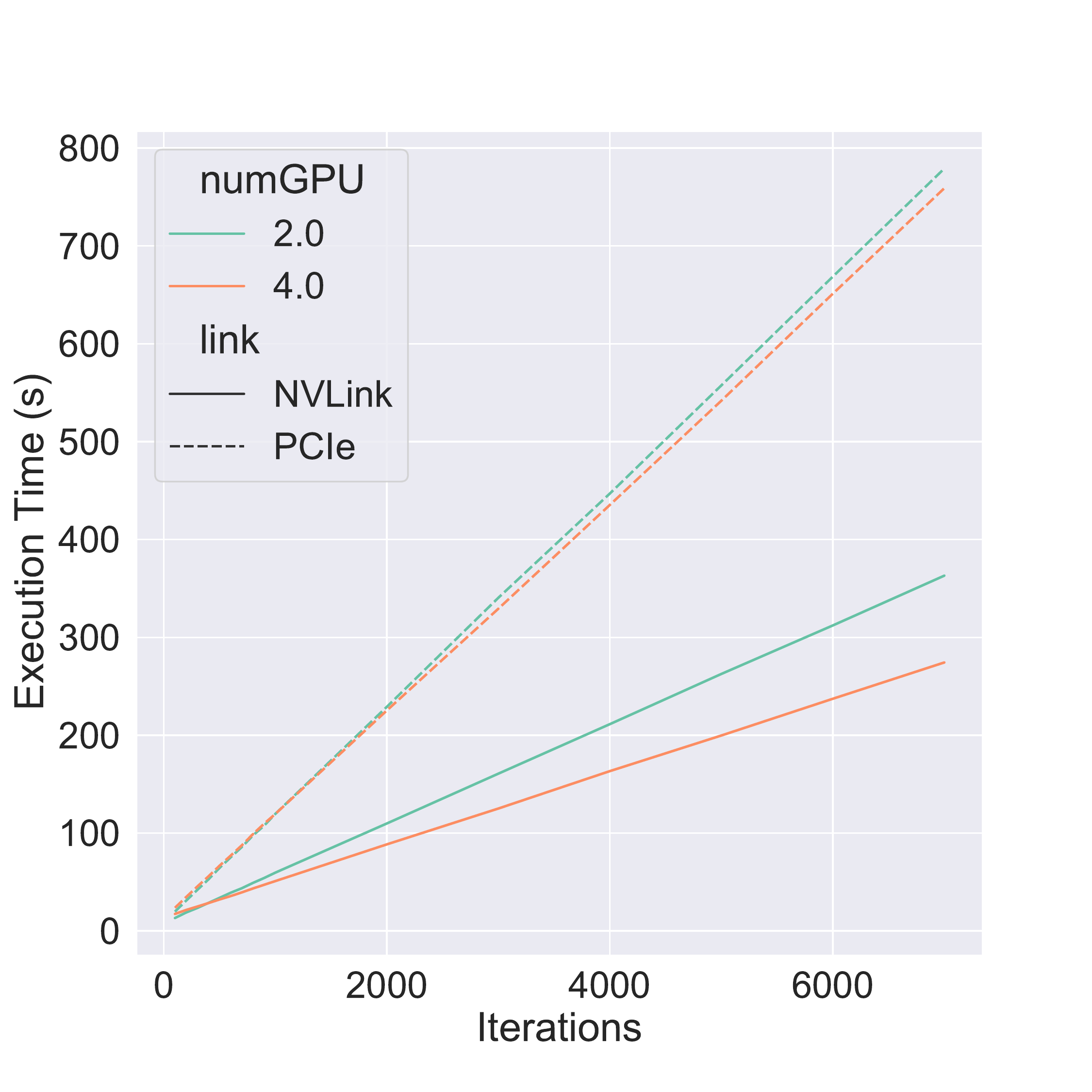}
        \vspace{-5mm}
        \caption{VGG-16 (Sensitive)}
        \label{fig:vgg_16_caffe_iter}
    \end{subfigure}
    \caption{Execution Time trends of Bandwidth Sensitive and Insensitive Networks.}
    \label{fig:exectime-bwsensitivity}
    \vspace{-3mm}
\end{figure} 

\begin{figure*}[!t]
    \includegraphics[page=1, scale=0.55]{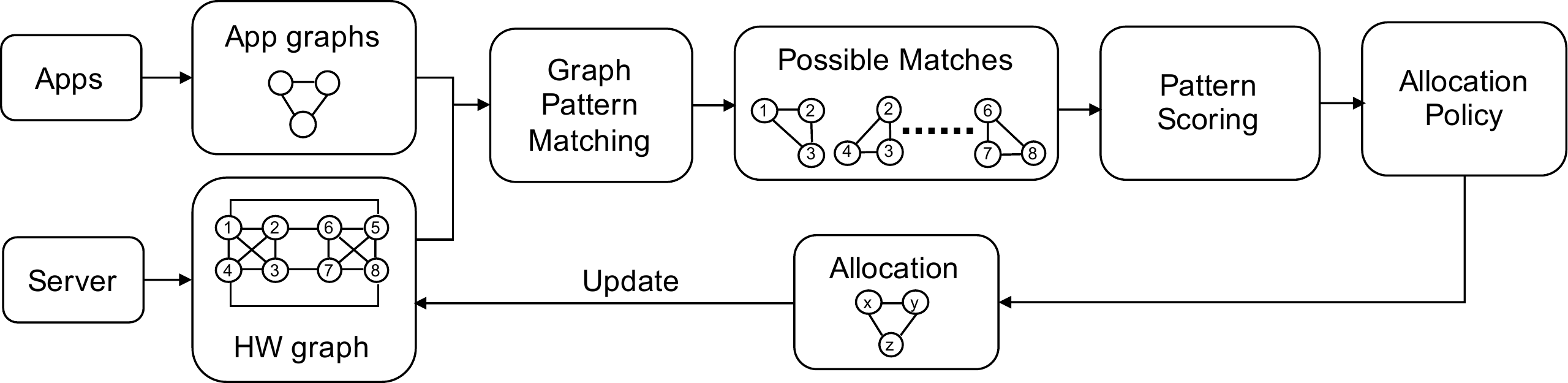}
    \caption{Overview of the \workfullname (\workname) system}
    \label{fig:overview}
    \vspace{-2mm}
\end{figure*}

We observe that a large majority of jobs receive suboptimal allocations. It should be noted that smaller jobs with less GPUs suffer more due to the potential for being spread out more across the interconnect topology. For example, with 3 GPU jobs, 75\% of jobs experience allocations with 20\% less bandwidth availability or worse and 25\% of jobs experience allocations with 45\% less bandwidth availability or worse. 

\subsection{Understanding Bandwidth sensitivity of ML workloads}
\label{subsec:understanding_bw_sensitivity}
As machine learning continues to spread across all aspects of modern life, it is no surprise that ML workloads are the most popular workloads for multi-accelerator systems~\cite{jeon2018multi, gandiva}. While these workloads are characterized to be very compute intensive, they have different degrees of sensitivity to the bandwidth provided by the system. 

Figure~\ref{fig:ml_comm_properties}(a) shows the distribution of data sizes that are communicated during the synchronization phase of ML training. Figure~\ref{fig:ml_comm_properties}(b) shows the number of collective communication calls per GPU that is employed in training these networks. We can infer from Figure~\ref{fig:ml_comm_properties}(a) that Alexnet, VGG, Inception, and CaffeNet involve an average communication data size of at least $10^5$ bytes during the synchronization. Similarly in Figure~\ref{fig:ml_comm_properties}(b), we can observe that Inception, Resnet, and GoogleNet involve a large number of communication calls.

It is also to be noted from Figure~\ref{fig:motivation_charts}(a) that data size has to be larger than $10^5$ bytes to make use of the available high-speed links. In GoogleNet, the number of communication calls are higher, however the average communication size is smaller than $10^5$ bytes. In CaffeNet, even though the average size is higher, there are not enough communication calls made to extract the benefit of high-speed links.

Hence, networks such as CaffeNet and GoogleNet are not bandwidth sensitive whereas VGG-16, Inception, Alexnet, and Resnet are. Furthermore, this assertion holds true when increasing number of GPUs and iterations as shown in Figure~\ref{fig:googlenet_caffe_iter} and Figure~\ref{fig:vgg_16_caffe_iter}. Other bandwidth sensitive networks such as Alexnet, Inception, and Resnet, and bandwidth insensitive networks, such as CaffeNet, follow similar trends to that of VGG and GoogleNet, respectively.

If a bandwidth sensitive network gets placed on a fragmented allocation, it may slowdown ML training jobs by more than 50\% as shown in Figure~\ref{fig:motivation_charts}(b). A solution that could potentially avoid the scenarios like this could improve overall throughput of the multi-accelerator systems.

In summary, the trends of heterogeneous link topologies and job co-location for multi-GPU servers can leave hardware resource fragmented.  Existing job allocation polices are unaware of the hardware diversity leading to a misappropriation of bandwidth to jobs.  Popular workloads such as ML training can be particularly susceptible to poor allocations.  Clearly, there is a need for a generalized allocation policy that can take into account the growing diversity of inter-accelerator interconnects and multi-accelerator workloads.  For the remainder of this work, we will focus our attention on GPUs and ML workloads, however our approach can be easily generalized to various accelerators and workloads.

\section{\workname: \workfullname}
\label{sec:Framework}

The \workfullname\ (\workname) framework introduces a generalized solution towards allocation of multi-accelerator workloads on multi-accelerator servers in multi-tenant (shared) environments such as cloud/enterprise data centers, virtualized environments, and shared high-performance computing facilities. Figure~\ref{fig:overview} shows an overview of \workname. 
Multi-accelerator applications and multi-accelerator servers are abstracted as smaller application graphs and larger hardware graphs, respectively. The application graphs capture the compute accelerator requirement and inter-accelerator communication topology of the workload, while the hardware graph captures the multi-accelerator system topology. 
In order to account for fragmentation and application bandwidth sensitivity, allocation decisions must consider the inter-accelerator communication properties of both the application and hardware. To solve this, we take a graph pattern matching approach where we mine the larger hardware graph (i.e., the data graph) for the smaller application graph (i.e., the pattern graph). Given a set of possible matches, we then assign a score to each pattern match to quantify the quality of each allocation and then select an allocation pattern using our proposed policy. In the remainder of this section, we will describe in detail each component of \workname.

\subsection{Application Topology}
\label{subsec:application_topology}
 
\begin{figure}[!t]
    \centering
    \includegraphics[width=0.5\columnwidth]{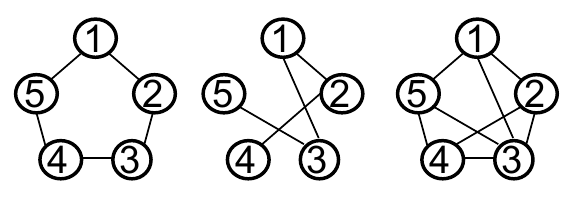}
    \caption{Example application topology for 5-GPU workload utilizing NCCL collective communication for inter-GPU communication. Application topologies can be ring (left), tree (middle), or a combination of both (right).}
    \label{fig:sample_app_topo}
\end{figure}

To make allocation decisions, \workname~ abstracts applications into \emph{application graphs} depicting the communication pattern across GPUs. In an application graph, vertices represent an accelerator compute resource (i.e. GPU) and the edges indicate communication between accelerators, as illustrated in Figure ~\ref{fig:sample_app_topo}. This application topology graph represents a summary of the application's communication pattern. While an application's communication pattern may vary over time, we cannot dynamically reallocate the hardware resource at runtime due to limited support for hardware preemption and the overhead of migration. Thus, we utilize a fixed application topology graph for allocation decisions.

Application communication patterns can be manually specified by the programmer, or can be automatically extracted through program analysis or profiling ~\cite{tallent2014palm,wilke2018compiler,faraji2016topology,ernsting2012algorithmic}. We will outline how each can be performed in the remainder of this subsection.

\begin{figure}[!t]
    \begin{subfigure}{\linewidth}
        \centering
        \includegraphics[width=\linewidth]{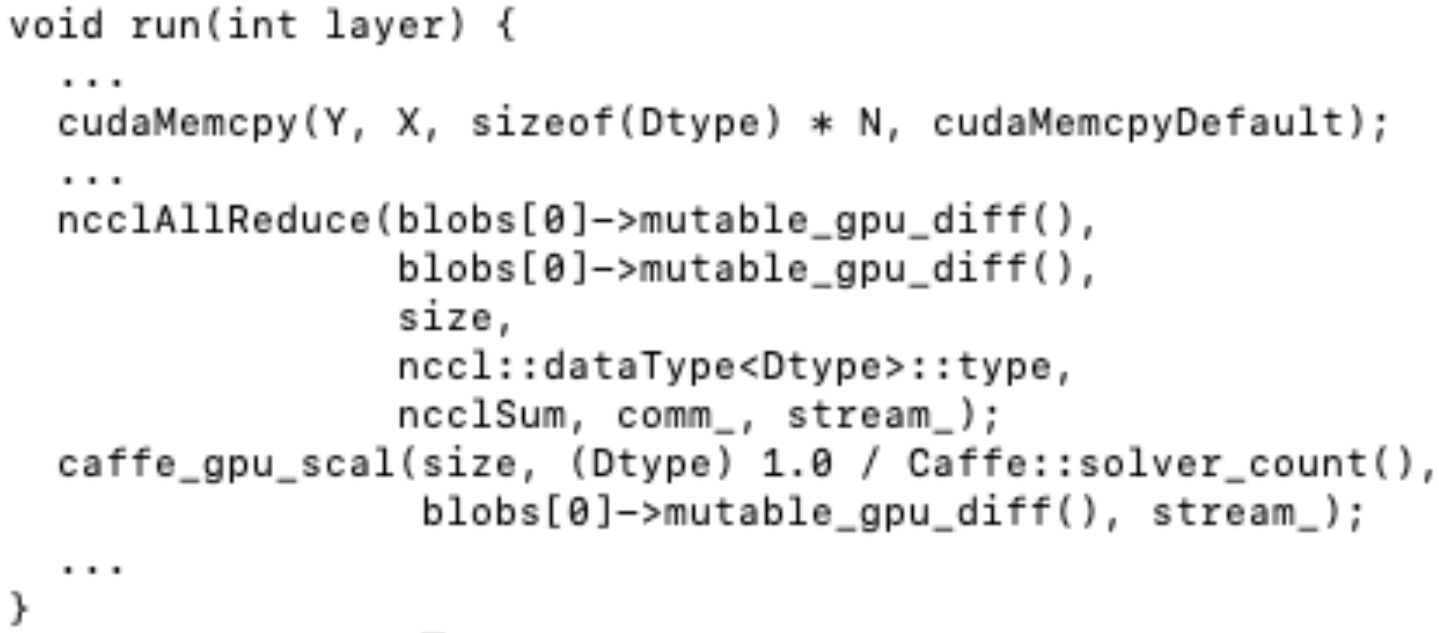}
        \caption{Sample multi-GPU CUDA program using NCCL.}
        \label{fig:sample_multi_gpu_program}
    \end{subfigure} %
    \begin{subfigure}{\linewidth}
        \centering
        \includegraphics[width=\linewidth]{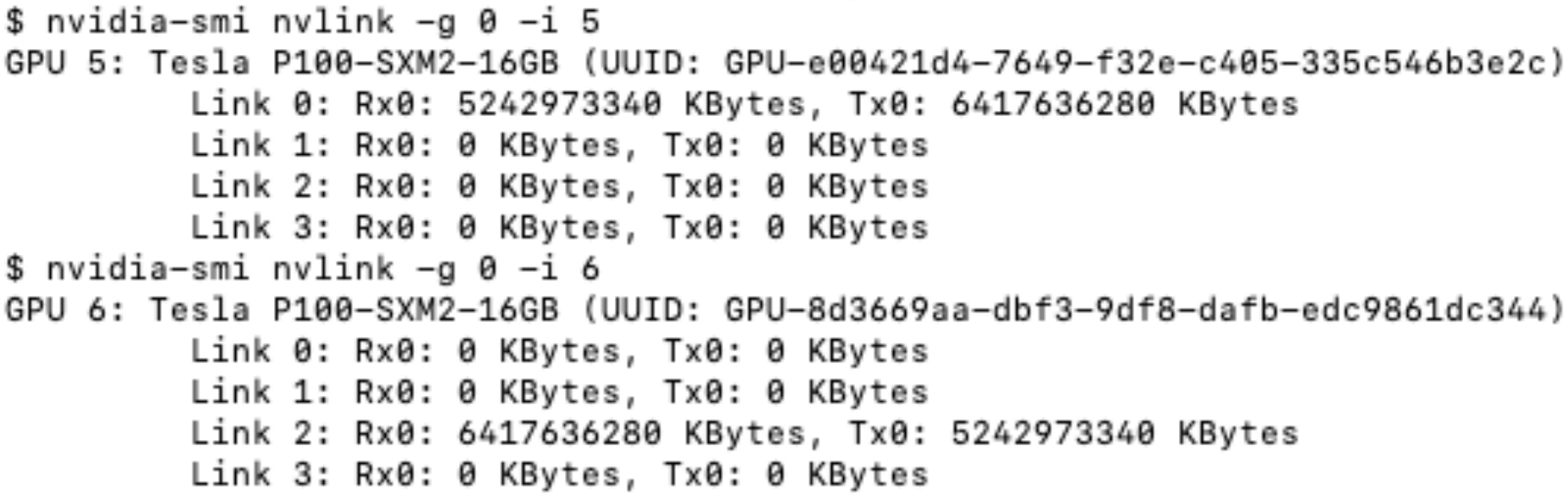}
        \vspace{-5mm}
        \caption{Sample NVlink traffic profiling.}
        \label{fig:nvlink_packets}
    \end{subfigure}
    \caption{Examples of identifying application topology through source code analysis and runtime profiling. }
    \label{fig:static_analysis_app_topo}
\end{figure}

\textbf{Source code analysis: }
Multi-GPU communication is typically coordinated through well-defined APIs.  Examples include the NCCL library for collective communications and \texttt{cudaMemcpyPeer()} (which explicitly passes the source and destination device) for peer-to-peer communication.  By identifying these API calls, communication patterns can be identified through a source code analysis. 
Figure~\ref{fig:sample_multi_gpu_program} illustrates this through a code sample from Caffe which performs the training operation of a layer. In this example, a collective all reduce is performed with \texttt{ncclAllReduce()} before the performing the layer's training computation in \texttt{caffe\_gpu\_scal()}.

NCCL handles collective communications by building rings or trees and utilizes them depending on the data transfer size that is required by the application.
Figure ~\ref{fig:sample_app_topo} shows potential application graphs for a 5-GPU allocation utilizing the NCCL library. Therefore, a 5-GPU application can have varying application topologies depending on the API that is used. Since the communication pattern can be identified based on the NCCL API, we can build an application topology graph by combining the graph of all NCCL API calls used in the program.

Besides NCCL and CUDA APIs, multi-GPU communication can also occur through MPI. For example, many HPC application pair a single MPI rank to a single GPU and use MPI calls to communication across ranks. With CUDA-aware MPI ~\cite{nccl-cuda-aware-mpi}, these GPU-GPU communication can be handled directly through NVLink without going through the host. While source code analysis of MPI calls can explicitly identify the communication pattern, many recent works have aimed to automatically identify MPI application topologies~\cite{faraji2016topology, ernsting2012algorithmic}, or automatically identify communication through compiler-assisted skeletons~\cite{tallent2014palm, wilke2018compiler}.

\textbf{Runtime profiling: }
Runtime profiling of multi-GPU workloads can identify an application's communication pattern through the monitoring of interconnect traffic over PCIe and NVLink. For example, tools such as \texttt{nvidia-smi} tracks the amount of traffic sent over each NVlink. Figure~\ref{fig:nvlink_packets} shows an example output for GPU 5 and 6. We can identify that these GPUs are directly connected by Link 0 of GPU 5 and Link 2 of GPU 6. Therefore, at runtime we can monitor the various interconnects to identify any inter-GPU communication between any given pair of GPUs to construct the application topology.

Runtime profiling is especially beneficial when a multi-GPU program has a complex and dynamic communication pattern that is implicit (i.e., Unified Memory) and cannot be easily identified through source code analysis. In these scenarios, instead of conservatively assuming a fully connected application topology, runtime profiling allows us to identify a more representative communication pattern enabling higher-quality allocations.

\subsection{Hardware Topology}
\label{subsec:hardware_topology}
In order to find an allocation, \workname~ aims to find a pattern (the application graph) in the larger graph representing the server hardware resource. 

In the hardware graph, the vertices represent the compute accelerators and edges are used to indicate the hardware links available on the server.  While the underlying system can have multiple paths (e.g. both an NVLink and PCIe) between two accelerators, edges are labeled with the highest available link bandwidth.  For simplicity, we assume the hardware graph to be fully connected graph as there always exists a path to each accelerator through the host.  For example, if two GPUs are directly connected with double NVLink-V2, then the edge will be labeled with 50. If two GPUs have no NVLink connectivity, then it will be labeled with the PCIe bandwidth of 12. 
The hardware graphs can be automatically extracted from existing tools, such as nvidia-smi, which describes how the accelerators and compute units are connected to each other. 

Note that our current approach only includes accelerators as vertices and not CPUs. We can potentially extend our approach to also include CPUs in both the application and hardware graph to account for CPU-GPU effects, such as potential NUMA effects. However, the goal of this work is to demonstrate the benefit of improving inter-accelerator communication and thus leave CPU-centric research for future explorations. Another challenge for the hardware topology representation is virtualized accelerators (e.g., Nvidia Multi-Instance GPU or AMD MxGPU) where jobs can be allocated to a virtual device and where inter-accelerator interconnects can be shared between multiple jobs. A potential solution to address this is to label the vertices (which represents a physical device) with the amount of physical resources available and then account for resource usage as resources are allocated to jobs and for the potential interference of the inter-accelerator interconnects. 

\subsection{Pattern Matching}
\label{subsec:pattern_matching}

To do the application to hardware graph pattern mining, we define a graph $g$ which contains a set of vertices $V(g)$ and labeled edges $E(g)$, a subgraph $s$ of $g$ which containing a subset of edges in $g$ and their endpoints. Given a hardware graph $G$ and the application pattern graph $P$, we aim to find a match $M$ which is a subgraph of $G$ that is isomorphic to $P$. Isomorphic is defined when there is a one-to-one mapping between the set of vertices in the application pattern graph $V(P)$ and the matching pattern graph $V(M)$ such that adjacent vertices in $P$ are also adjacent in $M$ with their corresponding vertices. 
This can be formulated as a \textit{subgraph isomorphism} (or subgraph matching) problem~\cite{Cordella_subgraph_matching_2004}. Several well-known algorithm exist in solving this problem, such as Ullmann's algorithm~\cite{Ullmann2011,Ullmann1976}, VF2~\cite{Cordella_subgraph_matching_2004}, and VF3~\cite{Carletti2018}. 
Since the goal of this paper is not in proposing a novel subgraph matching algorithm, we choose to utilize existing graph mining systems to implement \workname's pattern matching stage. Many general-purpose graph mining systems have been proposed, such as Arabesque~\cite{Arabesque}, AutoMine~\cite{AutoMine}, and Peregrine~\cite{Peregrine}. Specifically, Peregrine is a state-of-the-art fully pattern-aware graph mining system and pattern-aware programming model to create pattern-aware mining programs. Thus, we implement our pattern matching stage with Peregrine which takes our application pattern graph and hardware graph as input, and all matching subgraph patterns as outputs. 

This pattern matching scheme assumes one-to-one mapping between GPU applications and GPU hardware. Many-to-one mapping, where multiple applications can map to the same GPU hardware, are currently emerging. For example, GPUs can be virtualized for multi-tenancy~\cite{multi-tenant-cloud-scheduling} or GPUs can be hardware-partitioned into multiple GPUs (Nvidia multi-instance GPU). Identifying many-to-one mapping is non-trivial and is outside the scope of this work. However, \workname~ can potentially support many-to-one mapping by representing virtual GPUs as separate nodes in the hardware graph, or by labeling the nodes of the application / hardware graph with resource requirements / availability (threads, register, NVLink, etc.). This would require label-aware pattern matching and potentially partitioning of the application graph to fit into the available hardware resources, or utilize more complex many-to-one scheduling policies, such as in~\cite{multi-tenant-cloud-scheduling}.

\subsection{Pattern Scoring}
\label{subsec:pattern_scoring}
Given the set of matching patterns from the previous stage, \workname\ then must select the best pattern for allocation. \workname~ aims to assign a \textit{score} to each matching pattern which predicts which allocation will result in the most performance. To this end, we need to first answer \textit{How do we score each pattern match?} 

\begin{figure}[!t]
    \centering
    \includegraphics[width=0.85\linewidth]{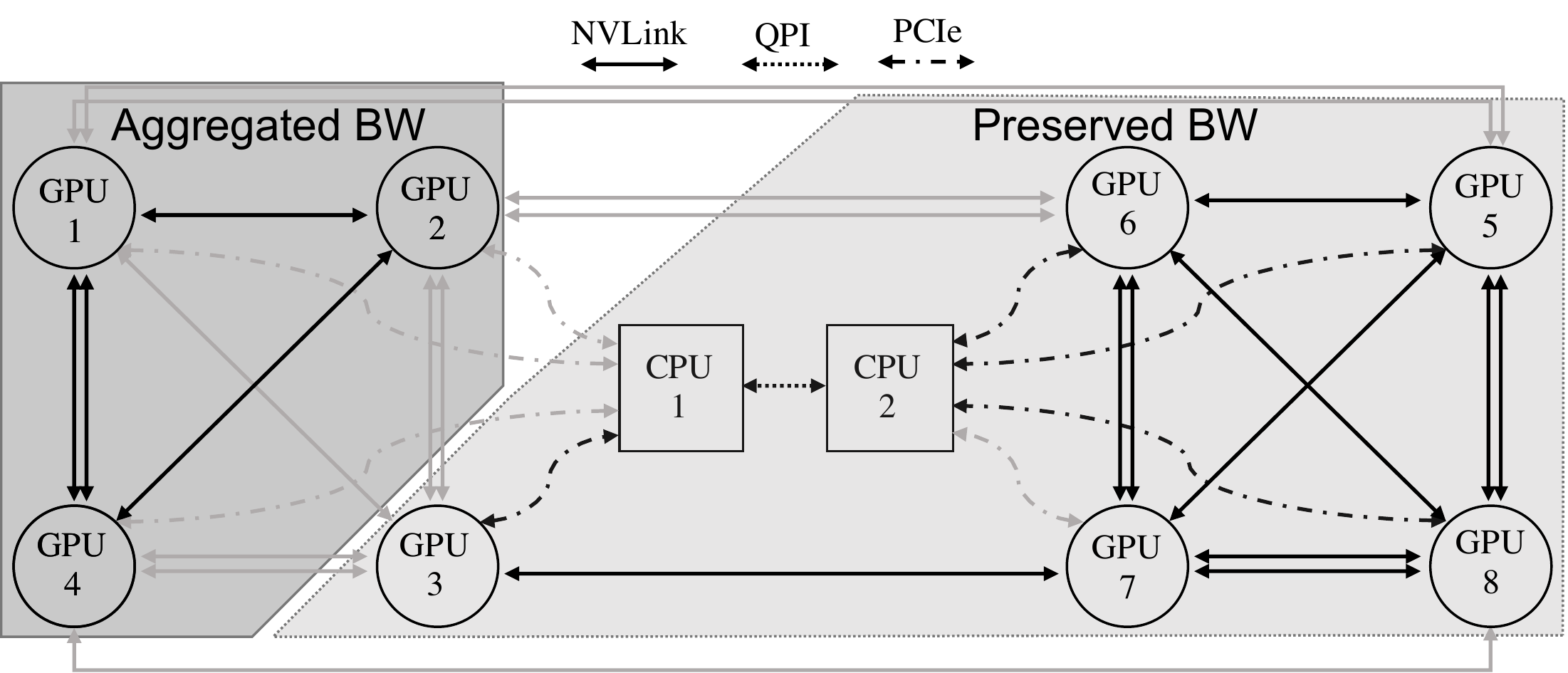}
    \caption{Illustrative example showing bandwidths used for Aggregated Bandwidth score calculation (left) and Preserved Bandwidth calculation (right) given an allocation [1, 2, 4]}
    \label{fig:agg_bw_illustration}
    \label{fig:preserved_bw_illustration}
\end{figure}

\begin{figure*}[!htb]
    \centering
    \begin{tikzpicture}
        \begin{groupplot}
            [
                group style =
                {
                group size = 3 by 1,
                horizontal sep = 1.5cm
                }
            ]
            \nextgroupplot[
                ylabel=$Exec.~Time~(s)$, xlabel=$Aggregated~BW~(GBps)$,
                width=0.6\columnwidth, height=3.5cm, ymin=0,
                legend style={legend columns=1, nodes={scale=0.7}},
                legend pos = south east, tick label style={font=\small},
                title style={at={(0.5,-0.55)},anchor=north,},
                enlarge x limits=0.15, bar width=0.3cm,
                title=(a) VGG-16 training Execution Time,
            ] \addplot [color=black,mark=*] table [x=LAST, y=execTime, col sep=space] {data/last-execTime.csv};
            
            \nextgroupplot[
                width = 0.8\columnwidth, height = 3.5cm,
                xlabel=$Aggregated~BW~(GBps)$, ylabel=$Eff.~BW~(GBps)$,
                ymin=0, tick label style={font=\small},
                title style={at={(0.5,-0.55)},anchor=north,},
                title=(b) Aggregated vs Effective Bandwidth,
            ]
            \addplot[only marks,] table [x=LAST, y=BW, col sep=space] {data/last-bw-data.csv};
            \nextgroupplot[
                ylabel=$Exec.~Time~(s)$, xlabel=$Effective~BW~(GBps)$,
                width=0.6\columnwidth, height=3.5cm,
                ymin=0,
                legend style={legend columns=1, nodes={scale=0.7}},
                legend pos = south east, tick label style={font=\small},
                title style={at={(0.5,-0.55)},anchor=north,},
                title=(c) VGG-16 training Execution Time,
            ] \addplot [color=black,mark=*] table [x=BW, y=VGG, col sep=space] {data/bw-execTime.csv};
        \end{groupplot}
    \end{tikzpicture}
    \caption{Evaluating pattern scoring metrics. (a) \textit{AggBW} does not correlate well with execution time. (b) This is due to \textit{AggBW} not correlating well with the effective achievable bandwidth of an allocation. (c) \textit{EffBW} correlates well with execution time. }
    \label{fig:aggregated_bw_allocations}
\end{figure*}
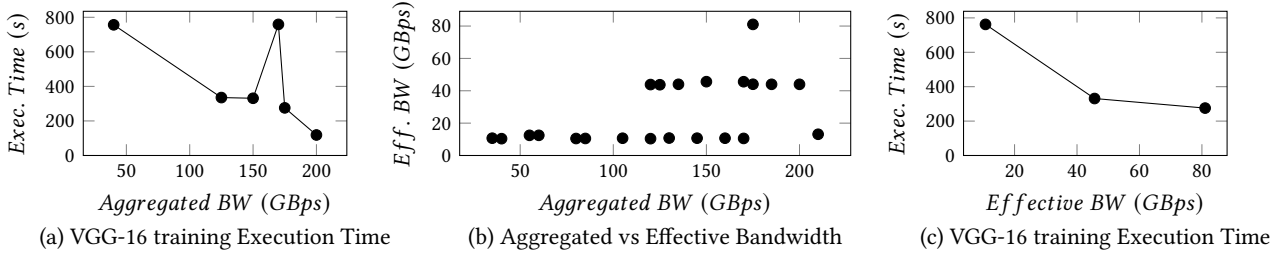

\subsubsection{Pattern scoring metrics}
\label{subsubsec:scoring_metrics}
To find a suitable pattern scoring metric, we explore two proposed metrics called \textit{Aggregated Bandwidth (AggBW)} and \textit{Effective Bandwidth (EffBW)}.

\textbf{Aggregated Bandwidth:}
We define \textit{Aggregated Bandwidth (AggBW)} as the total allocated bandwidth in the matching pattern $M$ that is used by the application pattern graph $P$. Since the application pattern graph $P$ is isomorphic to the matching pattern $M$, we know that $V(P) = V(M)$. However, the application's communication pattern may not use all of the available hardware interconnects that is allocated to it. That is, the set of edges in the application pattern may be a subset of the edges in the matching pattern, $E(P) \subseteq E(M)$. Therefore, the set of edges that are actually used by the application pattern in the matching pattern is denoted as $E(P) \cap E(M)$.  Recall that the edges $e$ of the hardware graph $E(G)$, and therefore the edges of the matching pattern $E(P)$, are weighted $w(e)$ with the highest available bandwidth between the two accelerator devices corresponding to the vertices of the graph.  Therefore, we formally define \textit{Aggregated Bandwidth} as shown in Equation~\ref{eq:aggregated_bandwidth}.

\begin{equation} \label{eq:aggregated_bandwidth}
    AggBW = \sum_{e \in \left( E(P) \cap E(M) \right)}{w(e)} ,
\end{equation}
 where $E(P) \cap E(M)$ represents the allocated interconnect in the matching pattern $M$ that are used by the application $P$, $e$ represents a used interconnect, and $w(e)$ represents the bandwidth of the interconnect.  
Specifically, \textit{AggBW} takes into account the application's communication pattern in order to quantify the amount of \textit{usable} communication bandwidth that was allocated to it. 

To illustrate \textit{AggBW}, Figure~\ref{fig:agg_bw_illustration} shows a possible allocation of a 3-GPU tasks that is mapped to GPU 1, 2, 4. Therefore, the \textit{AggBW} is the sum of the bandwidth of the interconnects between GPU 1,2 and 2,4 and 1,4. 

\textbf{Effective Bandwidth:}
We define \textit{Effective Bandwidth (EffBW)} as the peak achievable bandwidth for a given allocation. This metric is measured by running microbenchmarks to measure the peak effective real-world bandwidth across multiple links that is achievable for a given allocation. In our experiments, we use the NCCL All-reduce microbenchmark to determine the peak effective bandwidth. We selected this benchmark because the All-reduce collective communication pattern is the most used and has the greatest impact to overall execution time.
The effective bandwidth that we observe with different allocations is dependent on the number of links and the type of links (i.e. double NVLink, single NVLink, and/or PCIe).

\subsubsection{Evaluating Metrics } 
Now let us evaluate the two metrics, Aggregated and Effective Bandwidths.
We ran a multi-GPU training workload, VGG-16, with various 4-GPU and 5-GPU jobs and potential matching allocations. We measured the execution time of the workload, and the \textit{AggBW} and \textit{EffBW} of the allocation. 
Figure~\ref{fig:aggregated_bw_allocations}(a) shows that \textit{AggBW} does not correlate well with the workload's execution time. For example, an allocation with \textit{AggBW} of 170 is much slower than an allocation with \textit{AggBW} of 150. An ideal metric for scoring pattern matches would be correlated and be able to predict a workload's execution time.  

We find that this discrepancy is due to the fact that naively using the aggregated bandwidth \textit{AggBW} does not correlate with the effective bandwidth \textit{EffBW} that is achievable for a given allocation. This is demonstrated in Figure~\ref{fig:aggregated_bw_allocations}(b) which is collected using microbenchmarks to measure the effective bandwidth of various allocations ranging from 2-5 GPUs. \textit{Therefore, we find that execution time of workloads cannot be predicted by naively aggregating the allocated bandwidth. Instead, execution time of workloads must be predicted by the effective bandwidth.} 
Figure~\ref{fig:aggregated_bw_allocations}(c) demonstrates this fact by showing that effective bandwidth correlates well with workload execution time. 

However, a major challenge of using effective bandwidth as a metric to score matching patterns is that effective bandwidth cannot be trivially obtained given an allocation without microbenchmarking. Therefore, we need to create a model for predicting effective bandwidth.

\subsubsection{Predicted Effective Bandwidth}
\label{subsubsec:predicted_effective_bandwidth}

In the previous section, we demonstrated that the execution time is a function of effective bandwidth. Hence, we need to figure a way to predict \textit{EffBW} without having to run the microbenchmarks for a matching pattern. This could be achieved by solving a non-linear polynomial regression model. Here the Effective Bandwidth is related to the number of Double NVLinks ($x$), Single NVLinks ($y$), and PCIe links ($z$) in a given matching pattern $M$.

\begin{equation}\label{eq:non_linear_regression}
\begin{aligned}
    &Predicted~Effective~Bandwidth = \\&\theta_1 x + \theta_2 y + \theta_3 z +
                   \theta_4 \frac{1}{x+1} + \theta_5 \frac{1}{y+1} + \theta_6 \frac{1}{z+1} \\
                   &+ \theta_7 x y + \theta_8 y z + \theta_9 z x + \theta_{10} \frac{1}{xy+1} + \theta_{11} \frac{1}{yz+1} + \theta_{12} \frac{1}{zx+1}\\
                   &+ \theta_{13} x y z + \theta_{14} \frac{1}{xyz+1}
\end{aligned}
\end{equation}

To obtain data to train the model, we generate a set of 2, 3, 4, and 5-GPU allocations in a DGX-V machine described in Figure~\ref{fig:dgx_v100}. 
To limit the size of the generated set, we use an exhaustive set of allocations with unique $(x, y, z)$ resulting in a total of 31 samples. Next, we recorded the \textit{EffBW} by running the NCCL microbenchmark as described previously. Next, we solve equation~\ref{eq:non_linear_regression} using non-linear polynomial regression and the collected data (corresponding $(x, y, z)$ and the recorded \textit{EffBW}), to learn the relationships between the types of allocated links $(x, y, z)$ and \textit{EffBW}. 
Through the regression model in equation~\ref{eq:non_linear_regression}, we learn the coefficient $\theta$ of the following linear and non-linear features to capture their impact on effective bandwidth -- linear ($x$, $y$, $z$), inverse-linear ($\frac{1}{x+1}$,$\frac{1}{y+1}$,$\frac{1}{z+1}$), pairwise ($xy$, $yz$, $zx$), inverse-pairwise ($\frac{1}{xy+1}$,$\frac{1}{yz+1}$,$\frac{1}{zx+1}$), triplet ($xyz$), and inverse-triplet ($\frac{1}{xyz+1}$). The values of each of the coefficient is tabulated in table ~\ref{tab:weights_of_coefficients}.

\begin{table}[!htb]
    \small
    \centering
    \begin{tabular}{|c|c|c|c|c|c|c|c|} \hline
         \textbf{Coeff.} & \(\theta_1\) & \(\theta_2\) & \(\theta_3\) & \(\theta_4\) & \(\theta_5\) & \(\theta_6\) & \(\theta_7\) \\ \hline 
         \textbf{Value}    &	16.396 &	4.536 &	1.556  &	-20.694    &	-9.467    &	7.615    &	-7.973   \\ \hline	
         \textbf{Coeff.} &\(\theta_8\) & \(\theta_9\) & \(\theta_{10}\) & \(\theta_{11}\) & \(\theta_{12}\) & \(\theta_{13}\) & \(\theta_{14}\)  \\ \hline
         \textbf{Value} & 12.733    &	-4.195   &	-8.413   &	62.851    &	27.418    &	-5.114   &	-46.973   \\ \hline
    \end{tabular}
    \caption{Values of Coefficients.}
    \label{tab:weights_of_coefficients}
\end{table}

\begin{figure}[!h]
    \centering
    \begin{tikzpicture}
        \begin{axis}
        [
            xlabel=$Actual~Eff.~BW~(GBps)$,
            ylabel style={align=center}, ylabel=$Predicted$\\$Eff.~BW~(GBps)$, 
            width=\columnwidth, height=4cm, ymin=0, xmin=0,
            legend style={legend columns=1, nodes={scale=0.7}},
            legend pos = south east, tick label style={font=\small},
        ]
        \addplot [only marks, color=red,mark=*] table [x=BW, y=Predicted, col sep=space] {data/regression/2.csv};
        \addplot [only marks, color=blue,mark=otimes] table [x=BW, y=Predicted, col sep=space] {data/regression/3.csv};
        \addplot [only marks, color=brown,mark=triangle] table [x=BW, y=Predicted, col sep=space] {data/regression/4.csv};
        \addplot [only marks, color=yellow,mark=diamond] table [x=BW, y=Predicted, col sep=space] {data/regression/5.csv};
        \draw (axis cs:0,0) -- (axis cs:81.0265,68.70692109);
        \legend{2-GPU, 3-GPU, 4-GPU, 5-GPU}
        \end{axis}
    \end{tikzpicture}
    \caption{Predicted effective bandwidth correlates well with actual effective bandwidth and generalizes across jobs of different sizes.}
    \label{fig:regression_model}
\end{figure}
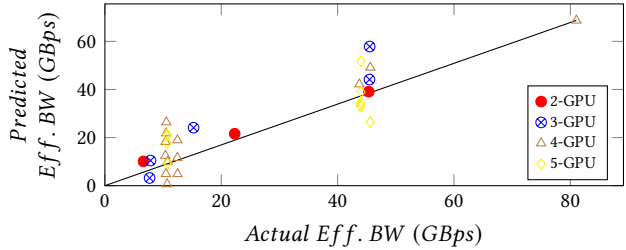

Figure~\ref{fig:regression_model} shows the predicted versus actual Effective Bandwidths given a $(x, y, z)$. 
For this model, the Relative Error, Root Mean Square Error (RMSE), and Mean Absolute Error (MAE) were found to be 0.0709, 1.5153, and 7.0539 respectively.
The model shows a strong correlation between the predicted \textit{EffBW} and the measured \textit{EffBW}, and generalizes well even when the number of GPUs in a job varies. This demonstrates that the Effective Bandwidth is strongly related to the mix of links allocated and not necessarily the amount of aggregate bandwidth allocated. Using equation~\ref{eq:non_linear_regression} we can now directly utilize \textit{EffBW} as our pattern scoring metric without the need for microbenchmarking.

\subsection{Pattern Selection Allocation Policy}
\label{subsec:pattern_selection}
Once the matching patterns are scored, \workname~ will then select a matching pattern for allocation. 
Recall form Section~\ref{subsec:understanding_bw_sensitivity} and Figure~\ref{table:bandwidth_sensitivity_matrix} that certain workloads are bandwidth sensitive while others are bandwidth insensitive. Thus, in order to maximize the overall performance of scheduled jobs, the pattern selection policy must account for (1) the effective bandwidth of an allocation, (2) the bandwidth sensitivity of the job, and (3) avoid starving future bandwidth sensitive jobs of effective bandwidth. To account for bandwidth sensitivity, \workname~assume that an application's bandwidth sensitivity is known and already annotated. The bandwidth sensitivity of an application can be determined through various means, for example, by profiling execution time vs allocated links as shown in Figure~\ref{fig:exectime-bwsensitivity}.

A novel aspect of \workname~is that when we select an allocation for bandwidth insensitive jobs, we try to \textit{preserve} as much remaining effective bandwidth as possible for future sensitive jobs. This bandwidth preservation scheme will then be able to optimize the allocation of bandwidth sensitive jobs. 
 In order to quantify the amount of remaining bandwidth that is preserved, we introduce a new metric as follows.

{
\setlength{\textfloatsep}{0.1cm}
\setlength{\floatsep}{0.1cm}
\begin{algorithm2e}[!b]
\small
    \SetAlgoLined
    \KwResult{Allocation}
        HWgraph hGraph\;
        AppGraph aGraph\;
        Allocation alloc = \{ \}\;
        Patterns possiblePatterns = graphPatternMatching (aGraph, hGraph)\;
        \If{aGraph is bwSensitive}{
            \ForEach(){pattern in possiblePatterns}{
                \If{EffectiveBW (pattern) > EffectiveBW (alloc)}{
                    alloc = pattern\;
                }{
                }
            }
        }\Else{
            \ForEach(){pattern in possiblePatterns}{
            \If{PreservedBW (pattern) > PreservedBW (alloc)}{
                alloc = pattern\;
            }{
            }
        }
        }
    \caption{Preserve Allocation Policy}
    \label{alg:preserve}
\end{algorithm2e}
}

\subsubsection{Preserved Bandwidth}
\label{subsubsec:preserved_bandwidth}
We define \textit{Preserved Bandwidth} as the aggregate bandwidth of the \textit{usable links that remain} (preserved) if a pattern match $M$ is allocated on the hardware graph $G$. The remaining hardware graph is denoted as $G \smallsetminus M$ which is the subgraph of $G$ \textit{induced} by the remaining available accelerator devices $V(G) \smallsetminus V(M)$. In other words, the remaining hardware graph $G \smallsetminus M$ is an \textit{induced subgraph} which is constructed by deleting the pattern match vertices $V(G) \smallsetminus V(M)$ (which allocates the corresponding accelerator devices) and with them all the incident edges (the hardware links that are no longer usable for future allocations). Figure ~\ref{fig:preserved_bw_illustration} illustrates the calculation of preserved bandwidth if GPUs 1, 2, and 4 are allocated.
Hence, we formally define \textit{Preserved Bandwidth} as follows in Equation ~\ref{eq:preserved_bandwidth}. 

\begin{equation}\label{eq:preserved_bandwidth}
    Preserved~Bandwidth = \sum_{e \in  E(G  \smallsetminus M) }{w(e)} 
\end{equation}

\subsubsection{Preserve Allocation Policy}
\label{subsec:preserve_policy}
We present the Preserve Allocation policy in Algorithm~\ref{alg:preserve}. 
In this policy, we rely on the programmer annotated bandwidth sensitivity ($bwSensitive$), the Preserved Bandwidth ($PreservedBW$) and Predicted Effective Bandwidth ($EffBW$). If the job to be allocated is bandwidth insensitive, we allocate the matching pattern that obtains the largest \textit{Preserved Bandwidth}. Meaning, we are preserving the amount of remaining available high-bandwidth links in the hardware graph for bandwidth sensitive allocations to avoid potentially starving these jobs. If the job to be allocated is bandwidth sensitive, we allocate the matching pattern with the highest \textit{Predicted Effective Bandwidth}.

\begin{figure*}[!ht]
    \begin{subfigure}[t]{0.5\textwidth}
        \centering
        \includegraphics[scale=0.2]{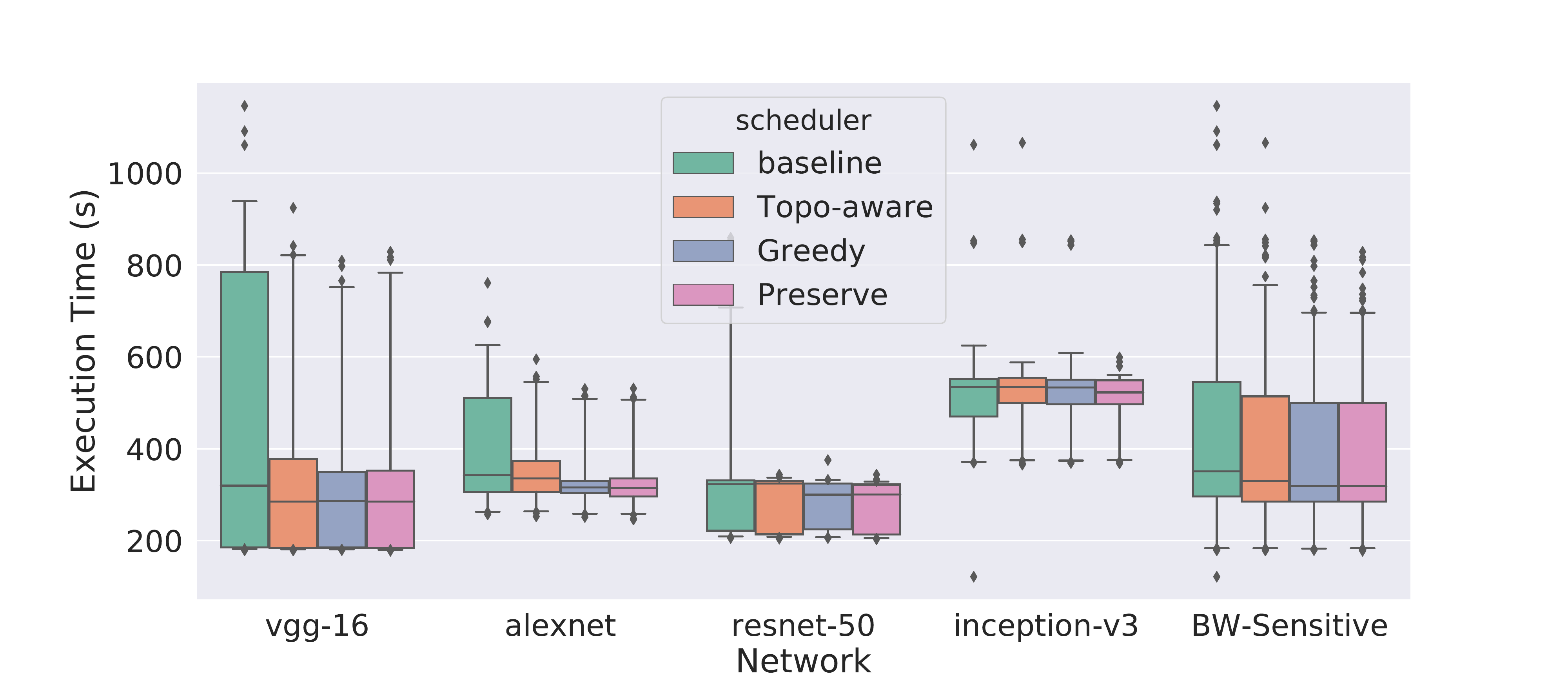}
        \caption{Execution time of bandwidth sensitive jobs }
    \end{subfigure}%
    \begin{subfigure}[t]{0.5\textwidth}
        \centering
        \includegraphics[scale=0.2]{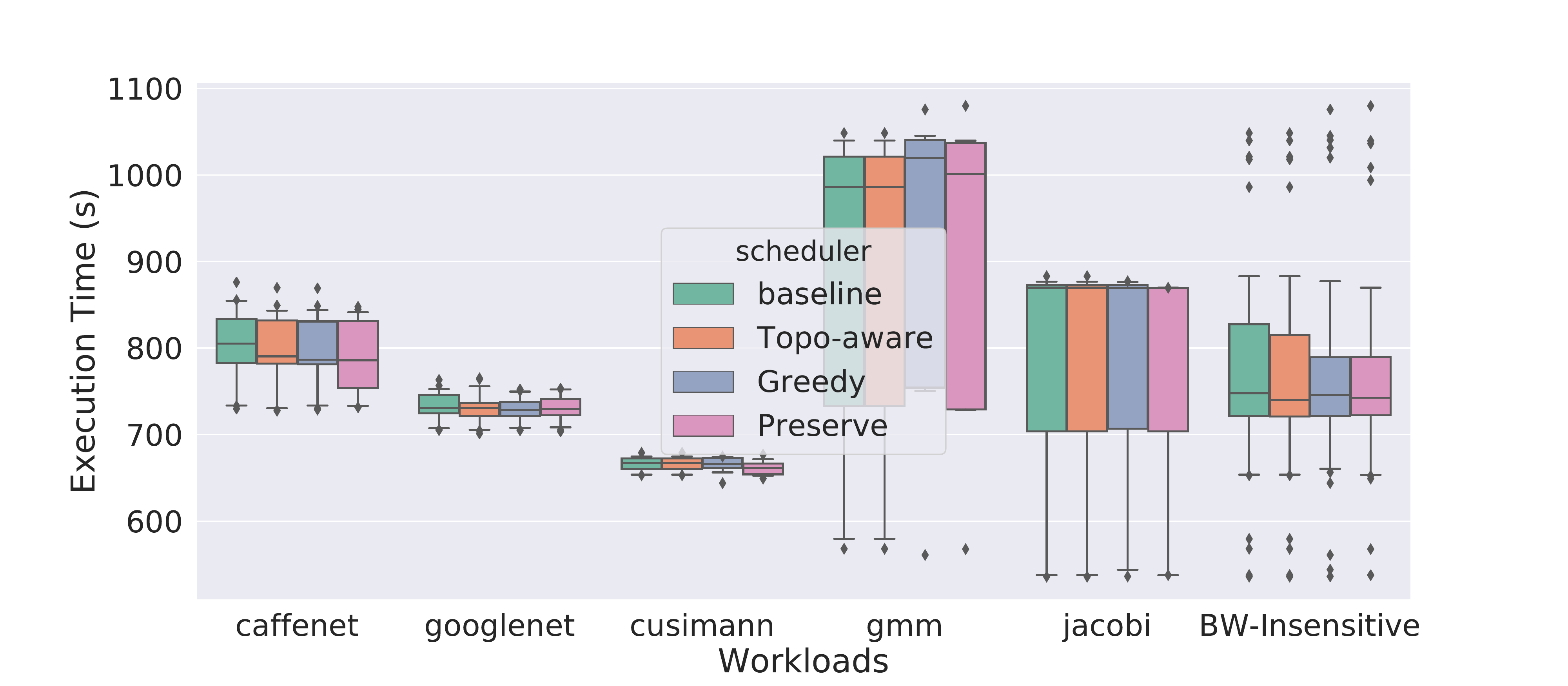}
        \caption{Execution time of bandwidth insensitive jobs}
    \end{subfigure}

    \begin{subfigure}[t]{0.5\textwidth}
        \centering
        \includegraphics[scale=0.2]{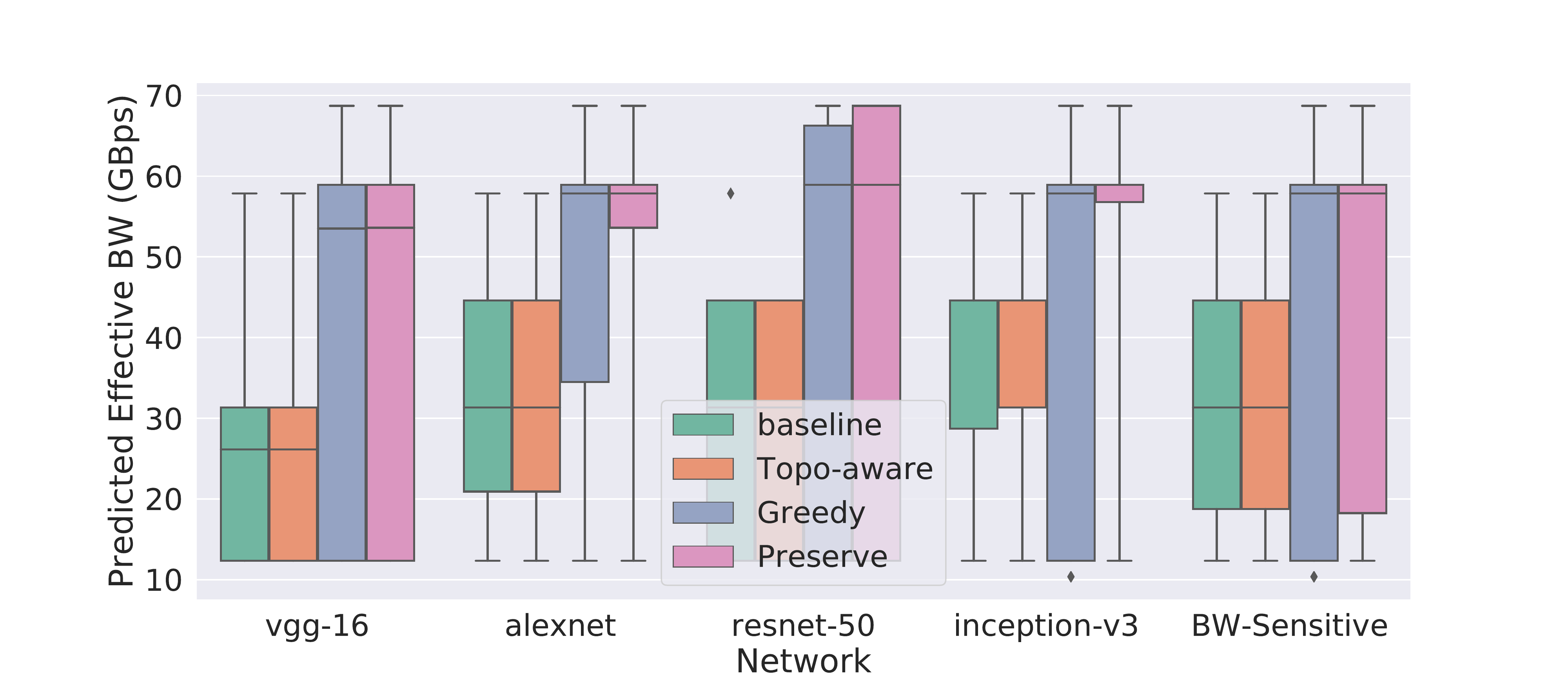}
        \caption{Effective bandwidth of bandwidth sensitive jobs}
    \end{subfigure}%
    \begin{subfigure}[t]{0.5\textwidth}
        \centering
        \includegraphics[scale=0.2]{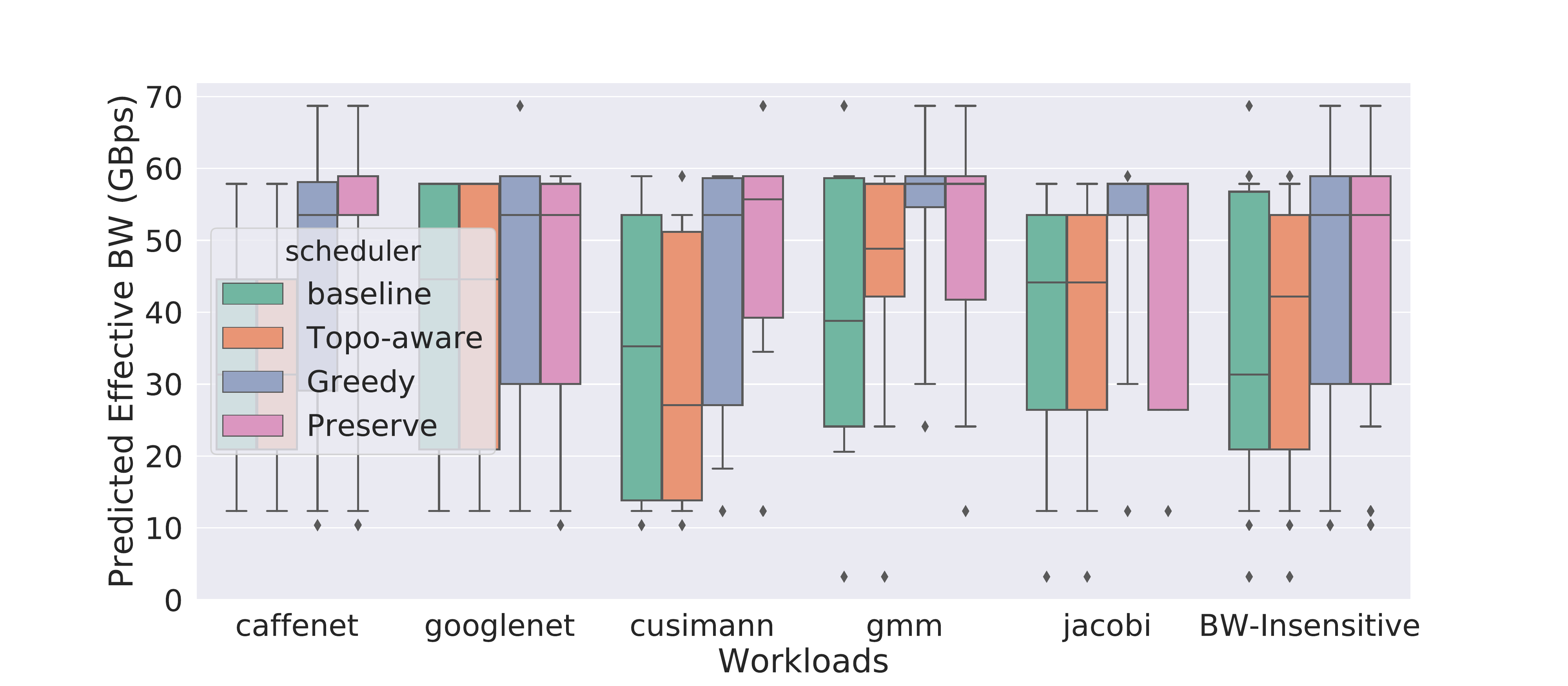}
        \caption{Effective bandwidth of bandwidth insensitive jobs}
    \end{subfigure}
    \vspace{-2mm}
    \caption{Evaluation results on DGX-V}
    \vspace{-3mm}
    \label{fig:real_run_results}
\end{figure*}

\subsection{State Management}
\label{subsec:state_management}
Once a matching pattern is selected for allocation, we then must update the hardware graph $G$. The hardware graph $G$ is updated whenever there is an allocation (a job is scheduled) and a deallocation (a job is finished). 
Once an allocation is obtained, we update the hardware graph to remove the unavailable vertices and incidental edges. When a job is complete and the hardware resource is deallocated, we update the hardware graph by adding back the vertices and incidental edges that was previously removed. 

\section{Evaluation}
\label{sec:evaluation}

To evaluate \workname, we use a combination of real-world runs and simulation. Specifically, we first evaluate the effectiveness of MAPA and the impact on performance on an NVIDIA DGX-1 V100 machine running on Ubuntu-16.04 with CUDA 11.3 and NCCL-2.10.3. The DGX-1 V100 hardware topology is shown in Figure~\ref{fig:dgx_v100}. The MAPA framework is built on top of Peregrine~\cite{Peregrine}, a graph mining engine, which performs subgraph pattern matching.
Although \workname~is agnostic to scheduling policies and can be extended to any scheduling policy and can employ reordering. However, in this work we use Fist-in First-out (FIFO) for scheduling jobs from the queue.

Later in Section~\ref{sec:evaluating_novel_hardware_topologies}, we will evaluate \workname~ on different multi-accelerator topology configurations by simulating the schedulers benefit on various representative hardware graphs. 

\textbf{Workloads: } In our evaluation, we use a set of Caffe~\cite{jia2014caffe} training jobs which makes use of multiple GPUs -- AlexNet~\cite{alexnet}, VGG-16~\cite{vgg}, Resnet-50~\cite{resnet}, Inception~\cite{inception}, GoogleNet~\cite{googlenet}, and CaffeNet~\cite{caffenet}.  
These neural networks are trained using the Image-Net dataset ~\cite{imagenet_cvpr09}. Each of the evaluated networks have different compute and communication patterns as discussed in Section~\ref{subsec:understanding_bw_sensitivity}. In addition, we use three other non-neural network multi-GPU workloads. They are a parallel simulated annealing algorithm for global optimization (Cusimann)~\cite{Tartan}, Gaussian Mixture Model (GMM)~\cite{Tartan}, and a Jacobi solver~\cite{jacobi-multi-gpu}. Furthermore, previous works ~\cite{Tartan, interconnectsgpus} have demonstrated that Cusimann and GMM to have negligible inter-GPU communication during the course of execution. Furthermore, we observed less than 3\% execution time improvement with Jacobi solver. Hence, we classify Cusimann, GMM, and Jacobi to be bandwidth insensitive. In this work, we focus on the inter-GPU communication aspect when multiple GPUs are employed in a single job.

\textbf{Jobs configuration: }
We randomly generated a job file of 300 jobs consisting of a uniform mix of training jobs for machine learning networks as shown in Figure~\ref{fig:ml_comm_properties}.

In addition, these jobs are generated with a random number of requested GPUs, from 1 to 5, which follows a uniform distribution. Prior work~\cite{jeon2018multi} has shown that the number of request GPUs for multi-GPU jobs in multi-tenant environments tend to be uniformly distributed. 

\textbf{Baseline Scheduling Policies: }
To evaluate \workname, we compare the preserve policy against three multi-GPU allocation policies---\textit{Baseline}, the current state-of-the-art scheduling technique \textit{Topo-aware}~\cite{amaral2017topology}, and a simple greedy policy \textit{Greedy}.
The \textit{Baseline} policy simply allocates GPU by ID by selecting the lowest IDs. This is how current GPU allocation are done in existing frameworks such as Nvidia Docker~\cite{nvidia_docker}. The \textit{Topo-aware} allocation policy~\cite{amaral2017topology} utilizes recursive bi-partitioning to select GPUs for allocation. This scheduler in effect selects GPU allocations under the same PCIe tree (CPU socket). The Greedy allocation policy simply selects a matching pattern with the highest \textit{Aggregated Bandwidth} for allocation.

\subsection{Evaluation on DGX-V System}
\label{subsec:evaluation_on_dgx_v}

We ran a mix of 300 jobs on the target DGX-1 V100 machine with Baseline, Topo-Aware, Greedy, and Preserve. These jobs are provisioned concurrently if sufficient hardware resources available to allow multiple jobs to run concurrently.
For each job we record the quality of the allocation using the predicted Effective Bandwidth score and the execution time. Figure~\ref{fig:real_run_results} shows our results, separated by sensitive and insensitive workloads. 

Figure~\ref{fig:real_run_results}(a) and ~\ref{fig:real_run_results}(b) shows the execution time of the experiments. 
Note that when running jobs on a multi-tenant server not all jobs will experience poor allocation due to fragmentation. Instead, the main point of focus should be the long tail of execution time where workloads that exhibit poor allocation will similarly exhibit poor execution time. 

The baseline policy allocates based on smallest available GPU ID and thus suffers significantly when allocations are fragmented, as demonstrated by the long tails of most bandwidth sensitive workloads, except Inception. The Topo-aware policy aims to schedule jobs under the same CPU socket which consists of fully inter-connected GPUs. This results in significantly improved tail execution times, most notably in VGG and Alexnet at the 75th percentile execution time, which improved from 785s to 378s and 511s to 374s, respectively. Overall, Topo-aware reduced the 75th percentile execution time from 540s to 505s for bandwidth sensitive jobs.  However, this Topo-aware policy is not generalized to support arbitrary application and hardware topologies. As shown in Figure~\ref{fig:real_run_results}(c) and (d), the chosen allocations' effective bandwidth does not significantly improve upon the baseline policy with the barplot of baseline and Topo-aware being nearly identical.

We evaluate \workname~with two pattern selection allocation policy--Greedy and Preserve. 
The \workname~Greedy policy greedily selects the allocation with the most aggregated bandwidth. Although aggregated bandwidth does not correlated with effective bandwidth, the Greedy policy nevertheless significantly improves the quality of allocation. As shown in Figure~\ref{fig:real_run_results}(c) and (d), the median effective bandwidth across all workloads (57.85GBps for Greedy and Preserve) is nearly the maximum effective bandwidth of baseline and Topo-aware which does not take into account application and hardware topologies. This demonstrates the benefits of \workname~ and the benefits of being application and hardware topology aware.

However, the Greedy policy does not consider application bandwidth sensitivity nor aim to preserve bandwidth for future bandwidth sensitive workloads. In  Figure~\ref{fig:real_run_results}(c) we see that the Greedy policy has allocations with lower 25th percentile of effective bandwidth (12.33 GBps), indicating that more sensitive jobs are starved.

\begin{table}[!ht]
    \small
    \centering
    \begin{tabular}{|c|c|c|c|c|c|c|}
        \hline
    	\textbf{Policy}       &   \textbf{MIN}	        & \textbf{25th \%}      & \textbf{50th \%}          & \textbf{75th \%}      &	\textbf{MAX}    & \textbf{Tput}     \\
    	\hline
        \textbf{Baseline}     &	 1.000                  &	1.000               &	1.000                   &	1.000               &	1.000           &   1.00               \\
        \hline
        \textbf{Topo-aware}   &   1.002                 &	1.029               &	1.385                   &	1.014               &	1.075           &   1.07           \\
        \hline
        \textbf{Greedy}       &	 0.997                  &	\textbf{1.059}      &	\textbf{1.519}          &	1.048               &	1.319           &   1.08           \\
        \hline
        \textbf{Preservation} &	 \textbf{1.006}         &	1.057               &	1.119                   &	\textbf{1.124}      &	\textbf{1.352}  &   \textbf{1.12}   \\
        \hline
    \end{tabular}
    \caption{Summary of results. Normalized execution time speed up and throughput (Tput) observed on DGX-1 V100.}
    \label{table:real_results_summary_exp}
\end{table}

\begin{figure*}[!t]
        \centering
        \includegraphics[width=0.8\linewidth]{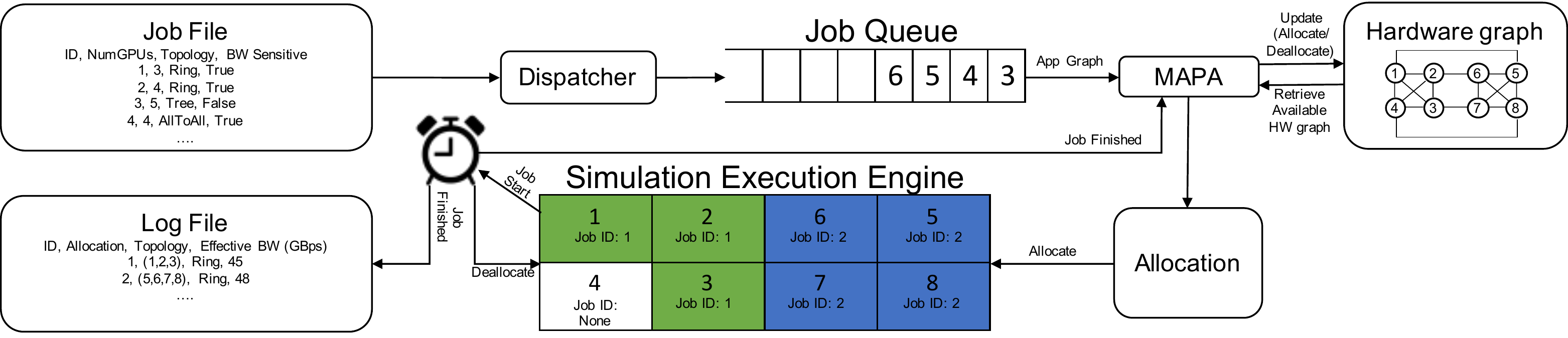}
        \vspace{-2mm}
        \caption{MAPA simulation execution framework.}
        \label{fig:simulation_execution_framework}
\end{figure*}

The Preserve policy is able to successfully preserve bandwidth for bandwidth sensitive workloads. This policy is able to achieve similar median effective bandwidth as the Greedy policy (57.85 GBps) 
without suffering at the 25th percentile. In many cases, the 25th percentile effective bandwidth also significantly improved as in the case of AlexNet and Inception-v3.  In terms of execution time, the Preserve policy achieves the lowest maximum tail execution time and 75th percentile execution time (498s) across the majority of the networks.

Table~\ref{table:real_results_summary_exp} summarizes the speedup across all allocation policies and the quartiles, normalized to the baseline policy. By greedily selecting the most aggregated bandwidth, the Greedy policy performs the best in the median case at the cost of less improvement for the longer running jobs at the tail. The Preserve policy is able to achieve the best speedup at the tail by improving the 75th percentile and Max by 12.4\% and 35.2\% over baseline. By improving the longer running jobs, the Preserve policy is able to improve throughput by 11.7\%. This throughput improvement is due to better utilization of available high-speed communication links, which results in higher GPU utilization and reduced execution times.

\section{Exploring Novel Hardware Topologies}
\label{sec:evaluating_novel_hardware_topologies}
\subsection{Methodology}
To explore the effects of scheduling and fragmentation for novel accelerator topologies, we built the \workname~ simulator framework to evaluate the quality of allocation for arbitrary hardware topologies. 

The simulation takes as input the hardware topology graph and a job file consisting of jobs represented by the application pattern graph and its execution times. For the job file input to the simulator, we obtained the extracted application pattern graph and measured baseline execution time from our real-world runs on the DGX-V. The output of the simulator is the effective bandwidths of each job. In lieu of building a full-featured performance model to predict the execution time of the workload, our simulator uses effective bandwidth as a proxy for execution time. 

\subsection{Simulation Framework}

Details of the simulation framework is shown in Figure ~\ref{fig:simulation_execution_framework}. 
The simulation starts with a job file. Each row in a job file corresponds to a job and is annotated with a job ID, number of GPUs, application topology, and bandwidth sensitivity. The \textit{Dispatcher} reads the job file and puts the job in the \textit{Job Queue}. The Job Queue employs a First-in First-out policy to mimic the FIFO scheduling in the real-world experiments. If there exist available GPU resources, the simulator invokes \workname~ to obtain an allocation for the next job.

The execution engine of the simulator is cycle-based and models the availability of a hardware resource. When a job is allocated, we flag the hardware as busy, record the cycle time, and begin the execution of the job. Once the specified execution time has elapsed, we flag the hardware resources as free, log the job's information into a log file, and send a \textit{Job Finished Signal} to ~\workname to update its hardware state. The logger records the Predicted Effective Bandwidth information along with other job properties.

\textbf{Simulator validation and soundness of effective bandwidth proxy. }
In order to validate the simulator with real-runs, we correlate the predicted Effective Bandwidth obtained in the real run results with the simulator configured for  DGX-V. As shown in Figure~\ref{fig:real_sim_effbw_scatter}, the simulated and real effective bandwidth correlates well indicating that the simulation adequately captures the scheduling behaviors of the real DGX-V system. 
We believe this simulation methodology can scale to evaluate future topologies since our evaluation metric (effective bandwidth) is based on the resource provisioned for a job, and not based on global topology properties. Therefore, we're confident our simulator result is accurate for future topologies utilizing the same link types.

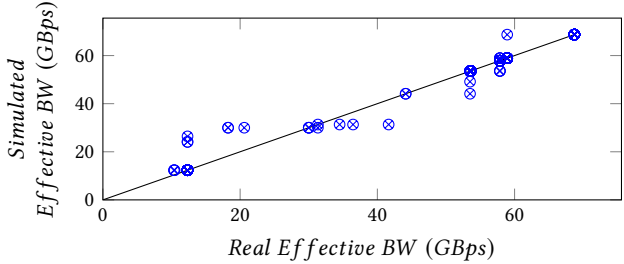
\begin{figure}[!t]
        \centering
        \begin{tikzpicture}[]
            \centering
            \begin{axis}
            [
                xlabel=$Real~Effective~BW~(GBps)$,
                ylabel style={align=center}, ylabel=$Simulated$\\$Effective~BW~(GBps)$,
                width=\linewidth, height=4cm, ymin=0, xmin=0,
                legend style={legend columns=1, nodes={scale=0.7}},
                legend pos = south east, tick label style={font=\small},
            ]
            \addplot [only marks, color=blue, mark=otimes] table [x=Real, y=Sim, col sep=space] {data/regression/sim_real_pred_bw.csv};
            \draw (axis cs:0,0) -- (axis cs:68.706921,68.706921);
            \end{axis}
        \end{tikzpicture}
    \caption{Effective bandwidth measured during DGX-V's real and simulation runs correlated well.}
    \label{fig:real_sim_effbw_scatter}
\end{figure}
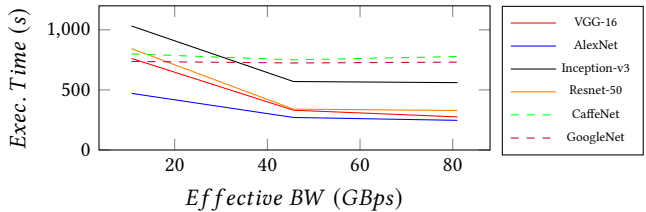
\begin{figure}[t]
    \centering
    \begin{tikzpicture}
        \begin{axis}
        [
            ylabel=$Exec.~Time~(s)$, xlabel=$Effective~BW~(GBps)$,
            width=0.8\linewidth, height=3.5cm, ymin=0, ymax=1200,
            legend pos = outer north east, legend style={legend columns = 1, font=\tiny},
            tick label style={font=\small},
        ]
        \addplot [color=red] table [x=BW, y=VGG, col sep=space] {data/bw-execTime.csv};
        \addplot [color=blue] table [x=BW, y=AlexNet, col sep=space] {data/bw-execTime.csv};
        \addplot [color=black] table [x=BW, y=Inception, col sep=space] {data/bw-execTime.csv};
        \addplot [color=orange] table [x=BW, y=Resnet, col sep=space] {data/bw-execTime.csv};
        \addplot [color=green,dashed] table [x=BW, y=Caffenet, col sep=space] {data/bw-execTime.csv};
        \addplot [color=purple,dashed] table [x=BW, y=Googlenet, col sep=space] {data/bw-execTime.csv};
        \legend{VGG-16, AlexNet, Inception-v3, Resnet-50, CaffeNet, GoogleNet}
        \end{axis}
    \end{tikzpicture}
    \caption{Effective bandwidth vs execution time observed during real runs on DGX-V.
    }
    \label{fig:effective_bandwidth_execution_time}
\end{figure}

To demonstrate the soundness of using effective bandwidth as a proxy for execution time, we collected the effective bandwidth and measured execution time of the real run for each workload. As shown in  Figure~\ref{fig:effective_bandwidth_execution_time}, we can see for bandwidth insensitive workloads that execution time is not impacted by effective bandwidth as expected. For bandwidth sensitive workloads, as effective bandwidth increases the execution time of the workload also improves (decreases). Although the amount of execution time improvement is limited once the effective bandwidth is past 50 GBps, the general trend still holds. Thus, effective bandwidth can be used as a good proxy for evaluating execution time improvements.  

\begin{figure}[!t]
    \centering
    \begin{subfigure}{0.5\columnwidth}
        \centering
        \includegraphics[width=0.9\linewidth]{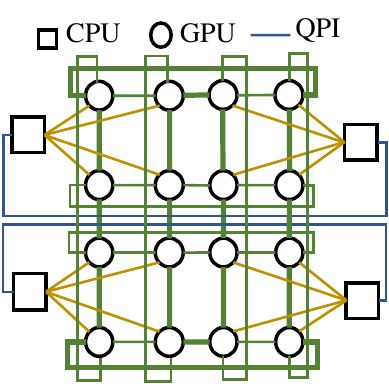}
        \caption{Torus-2d}
        \label{subfig:torus_2d}
    \end{subfigure}%
    \begin{subfigure}{0.5\columnwidth}
        \centering
        \includegraphics[width=0.9\linewidth]{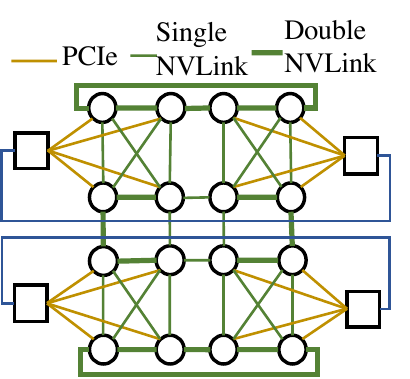}
        \caption{Cube-mesh}
        \label{subfig:cubemesh}
    \end{subfigure}
    \caption{16-GPU topologies}
    \label{fig:16-gpu_topologies}
    \vspace{-5mm}
\end{figure}

\textbf{Novel 16-GPU topologies.}
We explore the impact of scheduling policies on two novel 16-GPU hardware topologies -- Torus-2d, and Cube-mesh topologies.
The accelerators in Torus-2d and Cube-mesh topology are configured to have double NVLinks, single NVLinks, and PCIe as shown in Figures~\ref{subfig:torus_2d} and ~\ref{subfig:cubemesh}, respectively. Although 16-GPU topologies exists with crossbar switches (NVSwitch), we aim to explore alternative topologies consisting of cost-effective point-to-point links.

\subsection{Exploration Results}
 Recall that the aim is to improve the upper tail of execution time, and by proxy, to improve the lower tail (min and 25th percentile) of effective bandwidth. For brevity, we omit the results for bandwidth insensitive workloads since the execution times of these workloads are not impacted by effective bandwidth as shown in Figure~\ref{fig:effective_bandwidth_execution_time}.  
 
\begin{figure}[!t]
\vspace{-3mm}
    \begin{subfigure}{\linewidth}
        \centering
        \includegraphics[scale=0.23]{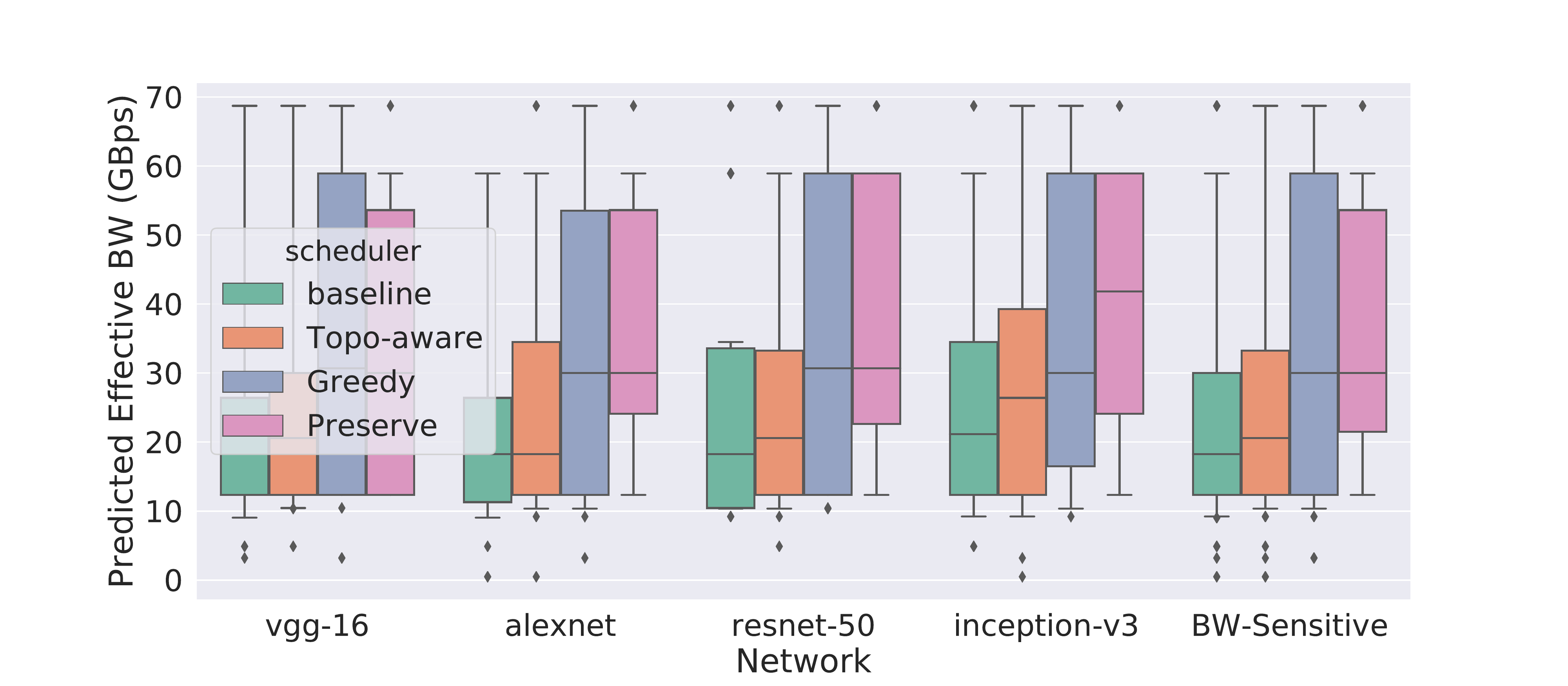}
        \caption{Torus-2d}
        \label{fig:torus_2d_simulation}
    \end{subfigure}

    \begin{subfigure}{\linewidth}
        \centering
        \includegraphics[scale=0.23]{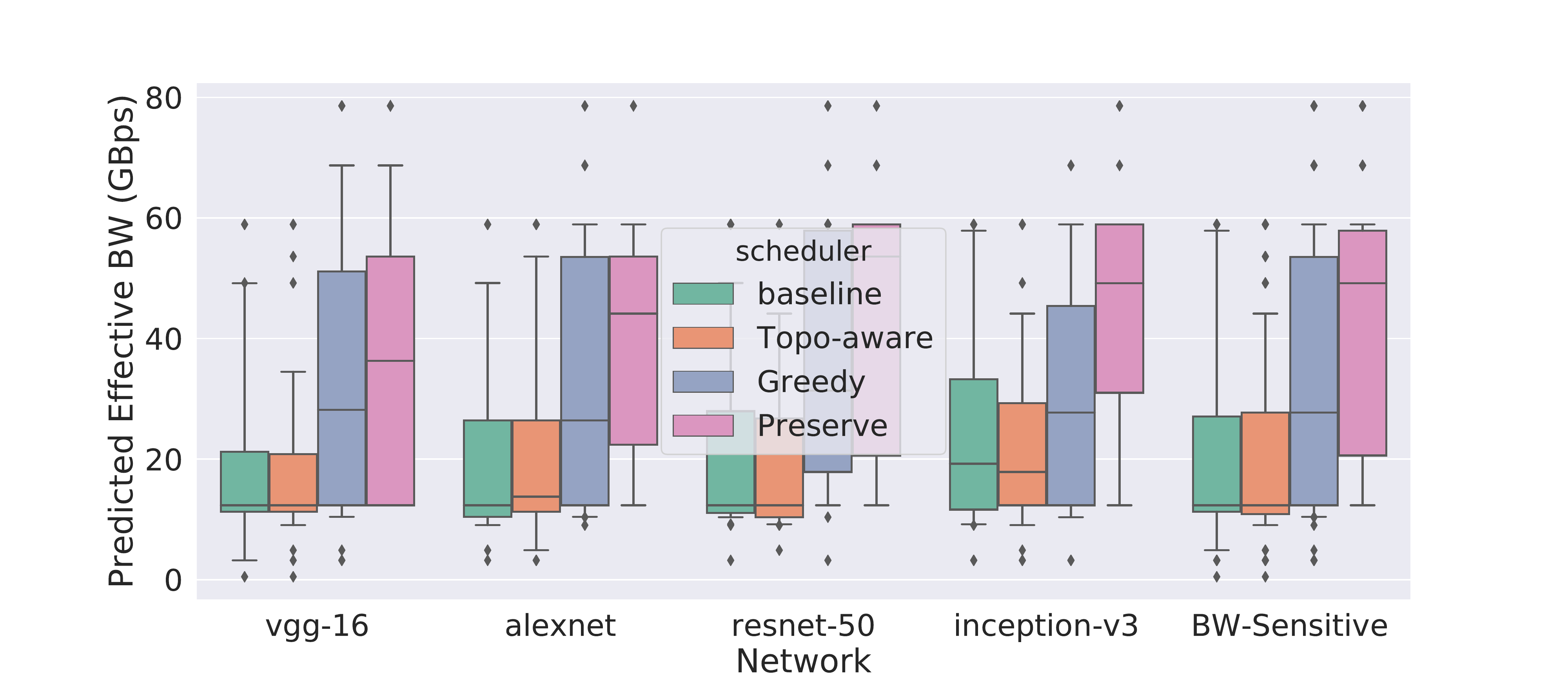}
        \caption{Cube Mesh}
        \label{fig:cubemesh_16_simulation}
    \end{subfigure}
    \caption{Simulation results for bandwidth sensitive workloads on 16-GPU topologies. Improvements to lower tail (min and 25th percentile) is better in both .  }
    \label{fig:16-GPU_simulation}
\end{figure}

 For the 16-GPU Torus-2d (Figures~\ref{fig:torus_2d_simulation}), we observe that Preserve significantly improves the 25th percentile and is better than the median of baseline and Topo-aware. In addition, the minimum of Preserve is equivalent to the 25th percentile of all other policies, demonstrating Preserve's ability to rein in the tail execution time. Due to the uniformness of the Torus-2d interconnect network, the Greedy policy is able to easily select high bandwidth allocations improving the 75th percentile (making fast jobs even faster).
 
 For the Cube-mesh topology (Figure~\ref{fig:cubemesh_16_simulation}), it is a more irregular network and thus more difficult to greedily select optimal allocations. Here, Preserve performs even better for sensitive workloads. While the minimum effective bandwidth of Preserve is equivalent to the 25th percentile of all other workloads, the median is near the 75th percentile of Greedy and the maximum of baseline and Topo-aware. \textbf{Therefore, half of the jobs allocated with Preserve will effectively run faster than the all of the jobs with baseline and Topo-aware and the majority of Greedy.}  
 
 These results demonstrates that as hardware topologies scale and becomes more complex and non-uniform, the greater the need for scheduling and allocation policies that are application communication pattern-aware and hardware topology-aware.

\subsection{Overhead of Scheduling}
\label{subsec:overhead_of_scheduling}
\begin{figure}[!t]
    \centering
    \begin{tikzpicture}
        \begin{axis}
        [
            mark size=1pt,
            ymode=log, ymin=1,
            xlabel=$Number~of~GPUs~Requested$, ylabel=$Overhead~(ms)$,
            legend pos=north west,
            xtick=data,
            width = \columnwidth, height = 3.4cm,
            title style={at={(0.5,-0.5)},anchor=north,},
            legend style={legend columns=4, nodes={scale=0.7}},
            enlarge x limits=0.1, bar width=0.15cm,
            tick label style={font=\small},
            ybar=0pt,
        ]
        \definecolor{color1}{RGB}{128,180,162}
        \addplot [fill=color1] table [ybar, x=Nodes, y=Summit, col sep=space] {data/policy-overhead.csv};
        \definecolor{color2}{RGB}{221,152,123}
        \addplot [fill=color2] table [ybar, x=Nodes, y=DGX-V, col sep=space] {data/policy-overhead.csv};
        \definecolor{color3}{RGB}{152,163,192}
        \addplot [fill=color3] table [ybar, x=Nodes, y=Torus-2d, col sep=space] {data/policy-overhead.csv};
        \definecolor{color4}{RGB}{209,152,190}
        \addplot [fill=color4] table [ybar, x=Nodes, y=CubeMesh-16, col sep=space] {data/policy-overhead.csv};
        \legend{Summit, DGX-V, Torus-2d, CubeMesh-16}
        \end{axis}
    \end{tikzpicture}
    \caption{Overhead analysis of \workname~ w/ Preserve policy.}
    \label{fig:preserve_policy_overhead}
\end{figure}
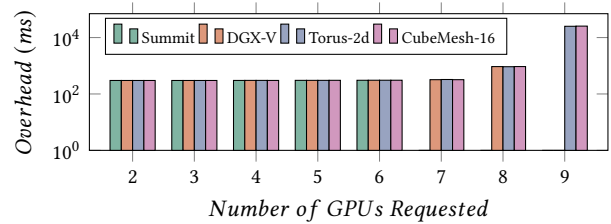

Figure~\ref{fig:preserve_policy_overhead} presents the scheduling overhead analysis of the \workname\ framework. We evaluate this overhead across different sizes of application pattern graphs (x-axis) and different sizes of hardware topology graphs. We evaluate hardware topology graphs of size 6, 8, and 16 for Summit, DGX-V and Torus-2d/CubeMesh-16, respectively. 
Typically, we observe scheduling overheads in the order of milliseconds which is negligible. However, the scheduling overhead does increase modestly for larger job sizes (9 GPUs and above) on larger hardware graphs (16 GPUs with 120$+$ edges). This is due to more combinations of matching patterns which requires more scoring of patterns. 

Note that this experiment is done on an idle hardware graph and \textit{represents an upper-bound of scheduling cost}. In reality, the allocation search will be performed on a smaller graph of available hardware which leads to significantly smaller pattern matches. Also our evaluation utilizes a single thread implementation to perform scoring. This overhead can be reduced by parallelizing the scoring process since it is a data parallel problem. Therefore, we expect our overhead to be manageable in real-world conditions and can scale to larger servers with parallel optimizations of our implementation.

\section{Related Works}
\label{sec:related_works}

\textbf{Scheduling for multi-node GPU clusters: }
Many works ~\cite{tripathy2021paver,abdolrashidi2017wireframe,tripathy2021localityguru,tripathy2020slumber,zamani2019greenmm,zamani2020saou,abdolrashidi2021blockmaestro,tan2016combating,li2019bstc,song2011iso,li2016critical,li2018warp,ductu2020independent,jahanshahi2020gpu-nest,warpapproximation,warpapprox,origami,warpedgates} have proposed optimizations to improve GPU performance and Energy efficiency.
Recent works, such as Gandiva~\cite{gandiva} and Philly~\cite{jeon2018multi}, proposes scheduling policies for multi-GPU jobs on multi-node multi-GPU clusters. Specifically, Gandiva proposes support for transparently migrating and time-slicing jobs for better job-to-GPU fit. Philly, on the other hand, aims to maximize the locality of multi-GPU allocations for non-preemptive multi-node multi-GPU clusters in multi-tenant environments. Both prior works aim to minimize fragmentation by either adding preemption support for migration, or by allocating across nodes to minimize fragmentation. Our work is complementary and aims to alleviate fragmentation that occurs \textit{within} the node itself in a multi-tenant environment due to the heterogeneity of links.

\textbf{Collective communication: }
In~\cite{wang2019blink, kobus2019gossip}, the authors have proposed techniques towards achieving efficient collective communication. Blink~\cite{wang2019blink} offers a new approach to collective communication by creating sets of spanning trees instead of rings. The spanning trees are dynamically generated based on the topology detected to utilize the links best. Specifically, given allocations from Philly, which are unaware of GPU-GPU interconnection topology, the goal of Blink is to identify optimal communication paths using spanning trees.  Gossip~\cite{kobus2019gossip} proposes flow-oriented collectives and generates transfer plans to best schedule packets. Works like WOTIR~\cite{ranganath2019speeding} presents software optimization techniques to improve the execution times of bad allocations using NVLink.  These works seek to optimize bad allocations, while our work seeks to reduce the number of bad allocations for bandwidth sensative jobs.

\textbf{Multi-GPUs for Machine Learning: }
From recent works ~\cite{gandiva, jeon2018multi, profilingdnndgx1},  Machine Learning (ML) is one of the primary workloads on multi-GPU systems. Hence, we use ML training as a target workload in this work, as well. We used Caffe~\cite{jia2014caffe} framework for Machine Learning in this work. These machine learning workloads use Nvidia Collective Communications Library (NCCL)~\cite{nccl} to perform operations like Reduce, AllReduce, Broadcast, Gather, Scatter, and Scatter-Gather. While we only demonstrated software NVLink routing in NCCL integrated into Caffe, our observed results and trends should generalize to other machine learning frameworks that use NCCL as the collective communication backend.
Besides, as ML models grow in size and complexity, the communication intensity will only increase, leading to a greater reliance on maximum achievable communication bandwidth.

\section{Conclusion}
\label{sec:conclusion_future_work}

In this work, we proposed ~\workfullname\\~(\workname), a novel approach to perform efficient scheduling and allocation of multi-accelerator workloads on multi-accelerator systems using a generalized graph pattern matching approach.
Through real-world evaluations, \workname~ improves overall system  throughput by up to 12\% and reduced the worst case execution time by 35\% over baseline. 
Through simulation we explore larger novel hardware topologies and find that \workname's benefit grow as hardware topologies scale and becomes more non-uniform. We demonstrate that more than half of the jobs allocated with \workname~ can effectively run faster than all jobs allocated with existing state-of-the-art scheduling policies. 

\begin{acks}
We thank the anonymous reviewers for their valuable feedback and suggestions. This work was partially supported by NSF grants \#1815643, \#1955650, \#2047521, and University of Sydney faculty startup funding and Australia Research Council (ARC) Discovery Project DP210101984. 

This work was also partially funded by the U.S. Dept. of Energy’s Office of Science Center for Advanced Technology Evaluation (CENATE) project under the Pacific Northwest National Laboratory. Pacific Northwest National Laboratory is operated by Battelle Memorial Institute for the U.S. Department of Energy under Contract DE-AC05-76RL01830.
\end{acks}

\bibliographystyle{ACM-Reference-Format}
\bibliography{refs.bib}

\end{document}